\documentclass[prb,aps,reprint,twoside,showpacs,preprintnumbers,amsmath,amssymb,citeautoscript]{revtex4-1}

\usepackage[dvips]{graphicx}% Include figure files
\usepackage{bm}% bold math

\usepackage[latin1]{inputenc}

\usepackage{hyperref}

\newcommand{\ka}{K}

\newcommand{\nn}[1]{\in \mathcal N_#1}

\newcommand{\br}{{\bf r}}
\newcommand{\bk}{{\bf k}}
\newcommand{\bv}{{\bf v}}

\newcommand{\bx}{{\bf x}}

\newcommand{\bA}{{\bf A}}
\newcommand{\bB}{{\bf B}}
\newcommand{\bE}{{\bf E}}

\newcommand{\bH}{{\bf H}}
\newcommand{\bJ}{{\bf J}}

\newcommand{\bM}{{\bf M}}
\newcommand{\bD}{{\bf D}}

\newcommand{\half}{\frac{1}{2}}

\newcommand{\df}{\nabla}
\newcommand{\db}{\overline\nabla}

\newcommand{\Eq}[1]{Eq.~(\ref{#1})}
\newcommand{\Fig}[1]{Fig.~\ref{#1}}

\newcommand{\av}[1]{\left<{#1}\right>}
\newcommand{\bigav}[1]{\big<{#1}\big>}

\newcommand{\vect}[1]{\mathbf{#1}}

\newcommand{\sumij}{\sum_{\av{ij}}}
\newcommand{\sumIJ}{\sum_{\av{IJ}}}
\newcommand{\In}{I_{ij}^n}
\newcommand{\Ic}{I_{ij}^C}
\newcommand{\Is}{I_{ij}^s}
\newcommand{\IR}{I_{ij}^R}
\newcommand{\IRn}{I_{ij}^{R+n}}
\newcommand{\Itot}{I_{ij}^\text{tot}}
\newcommand{\Itr}{I_{ij}^\text{tr}}

\newcommand{\tot}{\text{tot}}
\newcommand{\ext}{\text{ext}}
\newcommand{\BKT}{\text{BKT}}

\begin{document}

\title{%
Influence of vortices and phase fluctuations on thermoelectric transport properties
of superconductors in a magnetic field}

\author{Andreas Andersson}
\email{anan02@kth.se}

\author{Jack Lidmar}
\email{jlidmar@kth.se}

\affiliation{%
Theoretical Physics,
Royal Institute of Technology,
AlbaNova,
SE-106 91 Stockholm,
Sweden
}

\date{\today}

\begin{abstract}
  We study heat transport and thermoelectric effects in
  two-dimensional superconductors in a magnetic field.  These
  are modeled as granular Josephson-junction arrays, forming either regular or
  random lattices.  We employ two different models for the dynamics,
  relaxational model-A dynamics or resistively and capacitively
 shunted Josephson junction (RCSJ) dynamics.  We derive
  expressions for the heat current in these models, which are then
  used in numerical simulations to calculate the heat conductivity and
  the Nernst coefficient for different temperatures and magnetic
  fields.  At low temperatures and zero magnetic field the heat
  conductivity in the RCSJ model is calculated analytically from a
  spin wave approximation, and is seen to have an anomalous
  logarithmic dependence on the system size, and also to diverge in
  the completely overdamped limit $C \to 0$. From our simulations we
  find at low magnetic fields that the Nernst signal displays a
  characteristic ``tilted hill'' profile similar to experiments and a
  nonmonotonic temperature dependence of the heat conductivity.  We
  also investigate the effects of granularity and randomness, which
  become important for higher magnetic fields. In this regime
  geometric frustration strongly influences the results in both
  regular and random systems and leads to highly nontrivial magnetic
  field dependencies of the studied transport coefficients.
\end{abstract}


\maketitle

\section{Introduction}

Thermoelectric effects in superconductors are of considerable
interest, since they provide an important probe of fluctuations and
correlations in these materials.  Such effects have gained an
increasing amount of attention since the recent measurements of the
Nernst effect in the pseudogap phase of underdoped high-$T_c$
cuprates, where a remarkably large effect was
observed~\cite{Xu2000,*WangOng2006}.  The Nernst effect has since then
been measured in many other materials, e.g., in superconducting thin
films~\cite{Pourret2006,*PourretBehnia_NJP2009} and in the
iron-pnictides~\cite{Zhu2008}.  The Nernst effect is usually very
small for conventional metals, making it a particularly sensitive
probe of superconducting fluctuations.  Theoretical and numerical
studies have described the large Nernst effect either in terms of
superconducting fluctuations of Gaussian
nature~\cite{Ussishkin,MukerjeeHuse,Serbyn2009,Michaeli2009,*Michaeli2009a},
or as fluctuations of the phase of the order parameter, i.e., as a
result of vortices~\cite{WangOng2006,Podolsky,Raghu2008}.  Other
explanations of nonsuperconducting origin have also been put forward,
e.g., proximity to a quantum critical
point~\cite{Bhaseen2007,*Bhaseen2009,Sachdev},
quasiparticles~\cite{Oganesyan2004,Zhang2010}, stripe
order~\cite{Hackl2010}, etc.
Here we will focus exclusively on the effects of phase fluctuations
and vortices on heat and charge transport.  Phase fluctuations were
early on proposed to play a key role in the pseudogap phase of
underdoped cuprates~\cite{Emery1995}.  In quasi-two-dimensional
superconducting films and Josephson-junction arrays they are known to
be the dominant fluctuations~\cite{Berezinskii1,*Berezinskii2,*KT}.

In this paper we study the heat transport and the thermoelectric
response in two-dimensional granular superconductors or Josephson-
junction arrays, using two widely employed models for the dynamics,
(i) relaxational model-A dynamics, described by a Langevin equation,
or (ii) resistively and capacitively shunted Josephson junction (RCSJ)
dynamics. These are well suited for numerical simulation studies, and
have been used extensively to calculate electric and magnetic static
and dynamic properties, and to study vortex dynamics under influence
of electric currents~\cite{MonTeitel,ChungLeeStroud,Kim,Marconi}. The calculation
of thermoelectric properties is, however, less straightforward.
Previous simulation studies have used time-dependent Ginzburg-Landau
theory~\cite{MukerjeeHuse}, phase-only XY models~\cite{Podolsky} with
Langevin dynamics, or vortex dynamics~\cite{Raghu2008}.
They have been limited to rather narrow parameter regimes with a focus
on explaining the large Nernst effect seen in the pseudogap phase of
underdoped cuprates.
Here we present a comprehensive study of heat conductivity,
thermoelectric effects, and resistivity for a broad range of
parameters. We also investigate the effects of a granular structure.
The models are defined on a discrete lattice and can be experimentally
realized in granular superconductors and artificially fabricated
Josephson-junction arrays. At low magnetic fields discreteness effects
become less important so that our results in this regime are relevant
also for homogenous bulk superconductors.
On the other hand, in granular superconductors transport properties
are strongly affected by discreteness and geometric frustration
effects~\cite{Beloborodov2007}.  This is particularly true at high
magnetic fields when the inter-vortex distance becomes comparable to
the granularity. This leads to a rich structure in, e.g., the Nernst signal
as the magnetic field is varied, with anomalous sign changes occurring
in the vicinity of special commensurate
fillings~\cite{AnderssonLidmar}.

To start with, let us first recall that the heat current density $\bJ^Q$ and
electric current density $\bJ$ are related to the thermodynamic
forces,
the electric field $\bE$ and the temperature gradient $-\nabla T$,
by
the standard phenomenological linear relations
\begin{equation} \label{eq:thermoelectric1}
\begin{pmatrix}
\bJ^Q \\ 
\bJ
\end{pmatrix} = \begin{pmatrix}
\tilde\kappa & \tilde \alpha  \\ 
\alpha & \sigma 
\end{pmatrix}
\begin{pmatrix}
-\nabla T\\ 
\bE
\end{pmatrix} ,
\end{equation}
where the thermoelectric and electrothermal tensors obey the Onsager
relation $\tilde \alpha = T \alpha$.
The Nernst coefficient $\nu$ is defined as the off-diagonal response of the electric
field $E_y$ to an applied temperature gradient $\nabla_x T$ in a
transverse magnetic field $B_z$,
\begin{equation} \label{eq:nernst_def}
 \nu = \frac{e_N}{B_z} = \frac{1}{B_z} \frac{E_y}{(-\nabla_x T)} ,
\end{equation}
where $e_N$ is the so called Nernst signal. 
Both the Nernst effect and the heat conductivity are measured under
the condition of zero electric current, such that
\begin{align} \label{eq:nernst_alphasigma} 
  e_N &= \frac{\alpha_{xy}\sigma_{xx} - \alpha_{xx}\sigma_{xy}}{\sigma_{xx}^2+\sigma_{xy}^2}, \\
  \kappa &= \tilde\kappa - \tilde \alpha \sigma^{-1}\alpha .
\end{align}
In a system with no Hall effect ($\sigma_{xy} = 0$), which is the case
for the models studied below, \Eq{eq:nernst_alphasigma} reduces to
$e_N = \alpha_{xy} / \sigma_{xx}$.

In a phase-fluctuating superconductor, mobile vortices, either
thermally excited or induced by an applied magnetic field, may
significantly affect transport properties.
An applied electric current $\bJ$ will exert a perpendicular force on the
vortices and their motion will generate a transverse electric field
$\bE = -\bv \times \bB$ parallel to $\bJ$, thus causing a finite
flux-flow resistivity. Vortex motion can also be caused by a
temperature gradient, thereby generating an electric field
perpendicular to both the magnetic field and the temperature gradient.
The Nernst coefficient defined in \Eq{eq:nernst_def} can be seen
simply as the \emph{diagonal} response $\bv = -\nu\nabla T$ of the
vortex velocity to the temperature gradient.  For this reason it is
plausible that a large Nernst signal is expected in the vortex
liquid phase.

Via an Onsager relation a heat current can then be generated from an
applied electric current.  The vortices therefore also contribute to
the heat conductivity, although other heat carriers, e.g., phonons or
quasiparticles, probably dominate.
From the vortex point of view it is natural to consider the applied
current as the driving force and the electric field, which is
proportional (but perpendicular) to the \emph{vortex} current, as the
response. This yields an alternative, but equivalent formulation of
the linear relations in \Eq{eq:thermoelectric1}
\begin{equation}  \label{eq:thermoelectric2}
\begin{pmatrix}
\bJ^Q \\ 
\bE
\end{pmatrix} = \begin{pmatrix}
\kappa & \tilde \beta \\ 
\beta & \rho
\end{pmatrix}
\begin{pmatrix}
-\nabla T\\ 
\bJ
\end{pmatrix},
\end{equation}
where $\beta = - \tilde\beta/T = -\sigma^{-1}\alpha$.  This is the
approach employed in our simulations. Instead of calculating $e_N$
through other transport coefficients, i.e., using
\Eq{eq:nernst_alphasigma}, we obtain the Nernst signal directly as
$e_N = \beta_{yx} = \tilde \beta_{xy}/T$ for $\bJ = 0$.  The
longitudinal heat conductivity is just the diagonal components of the
tensor $\kappa$ in \Eq{eq:thermoelectric2}, and $\rho = \sigma^{-1}$
is the resistivity.

The picture described above applies when the vortices are free to move
in response to the driving forces. Pinning of vortices to material
defects, grain boundaries, etc., can lead to dramatic changes of the
transport properties.

The paper is organized as follows.
In Sec.~\ref{sec:models} we introduce the models and their dynamics,
Langevin or RCSJ, on general two-dimensional (2D) lattices.
In Sec.~\ref{sec:heat} we derive an expression for the heat current
for the models studied.  This has over the years proven to be a subtle
task, especially in the presence of a magnetic field. We present two
separate routes to finding the explicit form of the heat current for
Langevin and RCSJ dynamics. In addition we show, using a functional
integral approach, that the Nernst signal indeed can be calculated
from a Kubo formula involving the heat current.
Section~\ref{sec:linearized} discusses the thermal conductivity
at zero magnetic field in the low temperature limit, where a spin wave
approximation is applicable.  Our analytic calculations reveal a
logarithmic system size dependence of $\kappa$ in this regime for the
RCSJ case.
In Sec.~\ref{sec:methods} we give some technical details of our
numerical simulations, and
in the last part, Sec.~\ref{sec:results}, the results of our numerical
simulations on square and irregular lattices are presented. We
consider the case of zero and weak magnetic fields as well as the high
field limit, where the transport coefficients are severely affected by
geometric frustration. In the weak magnetic field limit the results
are discussed in relation to previous theoretical works and
experiments.

\section{Models}\label{sec:models}

We model a 2D granular superconductor (of size $L \times L$) as an
array of superconducting grains connected by Josephson
junctions. These grains may or may not be ordered and connected in a
regular fashion. The supercurrent flowing between two superconducting
grains $i$ and $j$ is given by the Josephson current-phase relation
\begin{align}								\label{eq:Is}
\Is &= I^c_{ij} \sin \gamma_{ij} ,\\
\gamma_{ij} &= \theta_i - \theta_j - \frac{2\pi}{\Phi_0} A_{ij},
\qquad
 A_{ij} = \int_{\br_i}^{\br_j} \bA \cdot d\br,
\end{align}
where $I_{ij}^c$ is the critical current of the junction,
$\Phi_0=h/2e$ is the superconducting flux quantum, and $\theta_i$ is
the superconducting phase of grain $i$.  The grains are assumed to be
small enough ($\lesssim$ the coherence length) that the phase is
constant over each grain.  Further, $\bA$ is the vector potential,
which we here separate into two parts
\begin{equation} \label{eq:A}
  \bA(\br,t) = \bA_{\text{ext}}(\br) + \frac{\Phi_0}{2\pi} \bm{\Delta}(t) ,
\end{equation}
where $\bA_{\text{ext}}(\br)$ is constant in time and corresponds to a
uniform magnetic field $\bB = \nabla \times \bA$ perpendicular to the
array, and $\bm{\Delta}(t) = (\Delta_x(t),\Delta_y(t))$ is spatially
uniform, but time dependent and is necessary to include in order to
describe temporal fluctuations in the average electric field $\bar \bE = -
\frac{\Phi_0}{2\pi} \dot{\bm{\Delta}}$, when periodic boundary conditions
are used~\cite{Kim,Marconi}.
Local fluctuations in the magnetic field $\bB$ and hence $\bA$ are ignored.

\subsection{Langevin dynamics}

The Langevin dynamics represents a phase-only description of the
time-dependent Ginzburg-Landau dynamics (TDGL) in uniform magnetic
field. The phenomenological equations of motion for $\{\theta\}$ and
$\bm{\Delta}$ are of local relaxation type
\begin{align}
\gamma \dot{\theta}_i &= -\frac{1}{\hbar} \frac{\partial H}{\partial \theta_i} + \eta_{i}
\\
\gamma_{\Delta} \dot{\bm{\Delta}} &= -\frac{1}{\hbar} \frac{\partial H} {\partial \bm{\Delta}} + \bm{\zeta} 
\\
H &= - \sumij J_{ij} \cos \gamma_{ij} , \label{eq:XY}
\end{align}
where the time constant $\gamma$ is dimensionless and $\gamma_\Delta =
\gamma L^2$, $H$ is the XY model Hamiltonian, and $\eta_{i}$ and $\bm{\zeta}$
are Gaussian stochastic white noise processes.

\begin{figure}[b]
\includegraphics[width = 6cm]{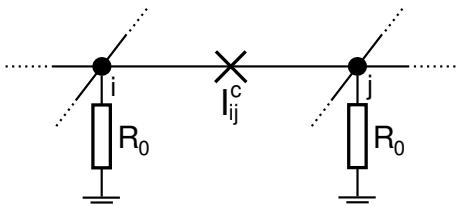}
\caption{Equivalent circuit model for Langevin dynamics. The cross
  denotes the Josephson junction with critical current $\Ic$.
\label{fig:circuit_langevin}
}
\end{figure}

These equations can be cast in the form of circuit equations for an
electric circuit built up using the elements displayed in
\Fig{fig:circuit_langevin}, where each site $i$ is
connected via a resistor $R_0 = \hbar/4 e^2 \gamma$ to ground:
\begin{align}
  \gamma \dot{\theta}_i &= \frac{V_i}{2e R_0} = - \frac{1}{2e} \sum_{j \nn i} I_{ij}^s + \eta_i \label{eq:langevin1} ,\\
 \gamma_\Delta \dot{\bm{\Delta}} &= \frac{1}{2e} \bigg( \sum_{\av{ij}} I_{ij}^s \vect{r_{ji}} - L^2 \bar{\vect{J}}^{\ext} \bigg) + \bm{\zeta}. \label{eq:langevin2} 
\end{align}
The sum in the first equation is taken over the set $\mathcal N_i$ of
superconducting grains connected to $i$ and is equivalent to a lattice
divergence. In the second equation $\av{ij}$ denotes a sum over all
links in the system. The vector $\vect{r}_{ji} = \vect{r}_j -
\vect{r}_i$ goes from site $i$ to site $j$ and $\bar{\vect{J}}^{\ext}$
is a fixed external current density.  
The Gaussian white noise terms $\eta_i$ and
$\bm{\zeta}$, which now can be interpreted as Johnson-Nyquist noise in
the resistors $R_0$, have the correlations
\begin{align}
\av{\eta_i(t)} = 0,& \quad \av{\eta_i(t)\eta_j(t')} = \frac{2k_BT\gamma}{\hbar} \delta_{ij}\delta(t-t') , \\
\av{\bm{\zeta}(t)} = 0,& \quad \av{\zeta_{\mu}(t) \zeta_{\nu}(t')} = \frac{2k_BT \gamma_\Delta}{\hbar}\delta_{\mu\nu}\delta(t-t') .
\end{align}

\subsection{RCSJ dynamics}

\begin{figure}[b]
\includegraphics[width = 7cm]{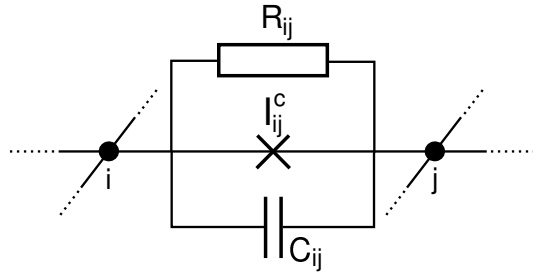}
\caption{Equivalent circuit model for RCSJ dynamics.
\label{fig:circuit_rcsj}
}
\end{figure}

In the RCSJ model each Josephson junction with critical current
$I^c_{ij}$ is shunted by both a resistor $R_{ij}$ and a capacitor
$C_{ij}$, see \Fig{fig:circuit_rcsj}. The RSJ model is obtained as a
special case when setting $C_{ij} = 0$.  We write the total current
from site $i$ to site $j$ as a sum of the super-, resistive,
capacitative, and noise currents
\begin{align} \label{eq:Itot_RCSJ}
I_{ij}^{\text{tot}} &=  I_{ij}^{s} + I_{ij}^{R} + I_{ij}^{C} + I_{ij}^{n} 
\nonumber \\
&=I_{ij}^c \sin \gamma_{ij} + \frac{V_{ij}}{R_{ij}} + C_{ij}\dot{V}_{ij} + I_{ij}^{n} ,
\end{align}
where the voltage between grain $i$ and $j$ is 
\begin{equation} \label{eq:voltage}
V_{ij}  \equiv V_i - V_j - \dot A_{ij} = \frac{\hbar}{2e}
\dot{\gamma}_{ij} ,
\end{equation}
and the Johnson-Nyquist noise in the resistors has zero mean
$\av{I_{ij}^{n}} = 0$ and covariance
\begin{equation} 
\av{I_{ij}^{n}(t) I_{kl}^{n}(t')} =  \frac{2 k_B T}{R_{ij}} 
(\delta_{ik}\delta_{jl} - \delta_{il}\delta_{jk}) \delta(t-t') .
\end{equation}

The equations of motion for the phases
$\{\theta_i\}$ and the twists $\bm{\Delta}$ are obtained from local
current conservation at each grain
\begin{align} 
\sum_{j \nn i} (I_{ij}^{\text{tot}} - \Itr) = 0. \label{eq:RCSJ1}
\end{align}
Here we introduced a ``transport current'' $\Itr$, which is a static,
divergence free current distribution, satisfying any external boundary
conditions, but is otherwise arbitrary.  One may, for instance,
connect some nodes to fixed external current sources or sinks. In the
present work we will usually use periodic boundary conditions instead,
together with the condition of a fixed average current density
$\bar\bJ^\text{ext}$ in the system,
\begin{align} 
\sum_{\av{ij}} I_{ij}^{\text{tot}} \vect{r}_{ji} &= \sum_{\av{ij}} \Itr \vect{r}_{ji} = L^2 \bar{\vect{J}}^{\text{ext}} .
\label{eq:RCSJ2}
\end{align}
For model purposes we may define a local current density on the links
of the lattice as
\begin{equation}
\vect{J}(\vect{r}) = \sum_{\av{ij}} \int_{\vect{r}_i}^{\vect{r}_j}
I^{\text{tot}}_{ij}\delta(\vect{r}-\vect{r'}) d\vect{r'} ,
\end{equation}
which directly leads to \Eq{eq:RCSJ2} when averaged over the system.

\subsection{Lattice structure}		\label{sec:lattice-structure}

\begin{figure}[h]
\includegraphics[width = 5cm]{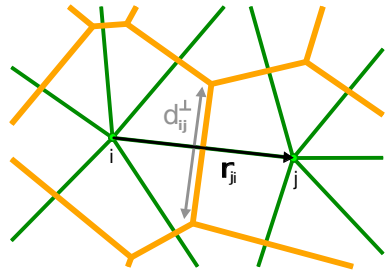}
\caption{(Color online) A part of a random lattice with the length $d^\perp_{ij}$
  and the vector $\br_{ji}$ defined. The green lines make up the direct Delaunay lattice and the 
  orange structure is the dual Voronoi lattice.
\label{fig:dperp_rji}
}
\end{figure}

We are interested in modeling both regular and random granular
superconductors.  At low magnetic fields the vortex separation is
large compared to the granules and the lattice structure does not
matter much.  In this regime the models approximate bulk
superconductors.  In the opposite limit of high magnetic fields the
lattice structure is important as frustration effects strongly
influence the transport properties.  Note that the formulation of the
models defined above is independent of lattice structure.
We will limit ourselves to two dimensions in the present work.
In our simulations presented below in Sec.~\ref{sec:results} we
consider not only square lattices, but also random lattices
appropriate as models of random granular superconductors.  These
irregular lattices are constructed by generating a set of randomly
distributed points $\br_i=(x_i,y_i)$ with unit density on a square and
connecting nearest neighbors by Delaunay triangulation. To control the
randomness we use the parameter $d_{\text{min}}$, which is the
shortest allowed distance between any two points in the system. Large
values of $d_{\text{min}}$ will thus create a relatively ordered
lattice structure, whereas lattices with small $d_{\text{min}}$ values
are more disordered.
For example, a given value of $d_\text{min} = \{0.0,0.4,0.6,0.8\}$
corresponds to a distribution of grain size diameters with standard
deviation $\{0.30,0.23,0.15,0.08\}$, respectively.
Examples can be seen in \Fig{fig:langevin_random_vs_f}.
For the random lattices we use two different models, one where the
critical current of every junction is set to a constant $I^c_{ij} =
I^c$ and one where the critical current is proportional to the contact
length $d^\perp_{ij}$ between the grains, $I^c_{ij} \sim d^\perp_{ij}$
(in the RCSJ case we also take $R^{-1}_{ij} \sim d^\perp_{ij}$ and
$C_{ij} \sim d^\perp_{ij}$).  This length is simply the distance
between the points in the dual Voronoi lattice, see
\Fig{fig:dperp_rji}.

\subsection{Transport coefficients}

In the models defined above the only nonzero transport coefficients
are the Nernst signal $e_N$, the diagonal thermal conductivity
$\kappa$ and the electrical resistivity $\rho$. These may be obtained
from equilibrium correlation functions using Kubo
formulas~\cite{Kubobook,Luttinger1964}.
While $\rho$ relates $\bE$ to a mechanical perturbation, $e_N$ and
$\kappa$ give the response to a nonmechanical property, namely a
temperature gradient, and the applicability of a standard Kubo formula
is not evident.  
Nevertheless, we show in the next section that the transport
coefficients can be expressed in standard form as
\begin{align} 
 \label{eq:nernst_kubo}
e_N &= \frac{\Omega}{k_B T^2} \int^{\infty}_{0} \bigav{\bar E_y(t) \bar{J}^Q_x(0)} dt , \\
 \label{eq:kappa_kubo}
 \kappa &= \frac{\Omega}{k_B T^2} \int^{\infty}_{0} \bigav{\bar{J}^Q_x(t)\bar{J}^Q_x(0)} dt , \\
 \label{eq:rho_kubo}
\rho &= \frac{\Omega}{k_B T} \int^{\infty}_{0} \bigav{\bar E_y(t) \bar E_y(0)} dt ,
\end{align}
where $\Omega = L_x L_y$ is the area of the system.
In these equations $\bar E_y = - (\Phi_0/2\pi) \dot \Delta_y$ is the
average electric field in the $y$-direction and
\begin{equation}								\label{eq:heat-current}
\bar J^Q_x = \frac{1}{\Omega} \sumij \left( x_{ji} \half (V_i + V_j) - x_{ij}^c \dot A_{ij}
\right) (\Itot - \Itr) 
\end{equation}
is the average heat current in the $x$-direction, with
$x_{ji}=x_j-x_i$ and $x_{ij}^c=(x_i+x_j)/2$.  
This form of the heat current, which is one of our main results, will 
be explicitly derived in the next section for the specific models 
under consideration.

\section{Heat current}\label{sec:heat}
In order to calculate thermoelectric effects and the heat conductivity
an expression for the heat current is needed.  Several microscopic
derivations of the heat current in superconductors has been given in
the literature~\cite{Schmid,Maki1968,CaroliMaki,Hu}. The presence of a
magnetic field yields a complication in that the total energy and
charge currents consist of magnetization currents in addition to the
transport currents~\cite{Obraztsov,CooperHalperinRuzin}.

Rather than relying on these microscopic expressions we derive here,
within the framework of the models defined in Sec.~\ref{sec:models},
an expression for the heat current that can be used consistently in
our calculations.  Below we will first derive the energy current by
considering the continuity equation for the energy density. This gives
the heat current after subtracting the nondissipative energy
transport, except for a magnetization contribution, which is
calculated separately in Sec.~\ref{sec:magnetization}.

Since we are interested in the response of the system to an applied
temperature gradient, which is a nonmechanical thermodynamic force,
the standard derivation of a Kubo formula does not hold.
One approach is to follow Luttinger~\cite{Luttinger1964}, and
introduce a ``gravitational'' field, which couples to the energy
density, and then proceed with the calculation of the response to
perturbations in that field.
For the present models, however, there is an alternative route.  With
the dynamics governed by stochastic equations,
Eqs.~(\ref{eq:langevin1}-\ref{eq:langevin2}) or
(\ref{eq:RCSJ1}-\ref{eq:RCSJ2}), in which the temperature enters via
the strength of the stochastic noise, it is possible to introduce
local temperature variations, such as a temperature gradient, and
calculate the resulting response.  
This calculation, done in Sec.~\ref{sec:Kubo} gives us both the Kubo formula
\Eq{eq:nernst_kubo} and the heat current \Eq{eq:heat-current}.
Note that in this setting, each point in the
model, or more precisely, each resistor in the circuit, is in contact
with a local heat bath.  For finite gradients one cannot expect the
heat current to be automatically conserved, since the resistors act as
sinks and sources.  For an infinitesimal thermal gradient the heat
current will, however, be conserved on average. Alternatively, one
could adjust the local temperatures to make sure that heat is
conserved on average also for finite temperature gradients. Such
self-consistent temperatures have been employed in studies of heat
conductivity in harmonic lattices~\cite{Bonetto2004}.  For the problem
at hand, however, the temperature profile is determined by the total
heat transport, including phonons, etc, so an externally imposed
temperature gradient is probably more realistic.  For the linear
response the form of the profile should not matter as long as it is
smooth.  We will use a linear temperature profile below.

\subsection{Heat current from continuity equations}
\label{sec:conserv}

In this section we will derive the heat current expressions for
granular superconductors described by Langevin and RCSJ dynamics, as
used in our simulations.  First, however,  it is useful to discuss
briefly the continuum formulation. 
Starting from the thermodynamic relation
\begin{equation}		\label{eq:Gibbs-Duhem}
 T ds = de - \mu dn - \bH \cdot d\bB - \bE \cdot d\bD
\end{equation} 
for a superconductor in a magnetic field $\bH$ and electric field $\bE$,
where $s$ and $e$ are the entropy and energy densities, $\mu$ the 
chemical potential, and $n$ the density of charge carriers (with 
charge $q$), one obtains for the heat current density
\begin{equation}						\label{eq:JQ_general}
 \bJ^Q = \bJ_\text{tot}^E - \frac{\mu}{q} \bJ - \bE \times \bH ,
\end{equation} 
where $\bJ$ is the electric transport current density.
The total energy current density is the sum of two parts, a
nonmagnetic part and Poynting's vector,
\begin{align}\label{eq:JEtot}
\bJ^E_{\text{tot}} = \bJ^E + \frac{1}{\mu_0} \bE \times \bB,
\end{align}
and the transport heat current density therefore also has two
contributions~\cite{LarkinVarlamov,Ussishkin,MukerjeeHuse,Hu}
\begin{align}
  \bJ^Q = \bJ^E - \frac{\mu}{q} \bJ + \bE \times \bM ,
\end{align}
where $\bM = \bB/\mu_0 - \bH$ is the magnetization.
The latter part can be rewritten as
\begin{align}								\label{eq:AxM}
  \bE \times \bM &= ( -\nabla \phi - \dot \bA ) \times \bM
  \notag \\
  &= - \nabla \times \phi \bM + \phi \bJ_M - \dot \bA \times \bM ,
\end{align}
where $\bJ_M = \bJ_\tot - \bJ = \nabla \times \bM$ is the
magnetization current, and $\phi$ the electric potential.  The first
term on the second line is purely rotational and will not contribute
to the heat transport when integrated over the system. The second term
$\phi (\bJ_\tot - \bJ)$ combined with the nonelectromagnetic part is
the standard expression $\bJ^E_\tot - (\xi/q) \bJ$, where $\xi = \mu +
q\phi$ is the electrochemical potential.  The last piece $-\dot\bA
\times \bM$ is with our gauge choice [\Eq{eq:A}] spatially constant.
Therefore it cannot be uniquely determined via the continuity
equations below, but will have to be added separately later.

Note the dual role played by the subtraction $\bE \times \bH$ in
\Eq{eq:JQ_general}. It contains the subtraction $\phi \bJ$
(but in a gauge invariant way).  At the same time it subtracts the
magnetic field energy transported by the \emph{vortex} current, since
$H_z \Phi_0$ can be viewed as a magnetic contribution to the
\emph{vortex} chemical potential, while the electric field $\bE$ is
proportional (but transverse) to the vortex current $\bJ_v = \bE
\times \hat {\mathbf z}/\Phi_0$.
In principle one should also subtract the nonelectromagnetic energy
transported by vortices $\mu_v \bJ_v$.
In the present models, however, $\mu_v = 0$. In fact, the chemical
potential $\mu$ for the charge carriers is also zero.
We will now derive the analogous expressions for the discrete models.

\subsubsection{Langevin dynamics}
It is instructive to study a slightly more general model which
includes the charging energy of the grains, described by a circuit
model with capacitors $C_0$ added in parallel with the resistors $R_0$
to ground.  This modification, besides being more general, makes the
derivation more physically transparent, while leaving the final
results unchanged.  The total energy for this model can be written
\begin{equation}
E = -\sum_{\av{ij}} J_{ij} \cos \gamma_{ij} + \sum_i \frac{1}{2}C_0 V_i^2 ,
\end{equation}
with $\gamma_{ij} = \theta_i - \theta_j - \frac{2\pi}{\Phi_0} A_{ij}$ being the gauge
invariant phase difference between site $i$ and $j$ and $V_i = \hbar
\dot{\theta}_i / 2e$ the voltage at site $i$ to ground.
With the site $i$ we associate a local energy
\begin{equation}
e_i = -\frac{1}{2} \sum_{j \nn{i}} J_{ij} \cos \gamma_{ij} +
\frac{1}{2}C_0 V_i^2 .
\end{equation}
The time derivative of this is
\begin{equation} \label{eq:edot_langevin}
\dot{e}_i =  \frac{1}{2} \sum_{j \nn{i}}  J_{ij} \dot{\gamma}_{ij} \sin
\gamma_{ij} + C_0 V_i \dot{V}_i .
\end{equation}
In the first term we may identify $J_{ij} \dot \gamma_{ij} \sin \gamma_{ij} = V_{ij} \Is$.
The last term contains the current through the capacitor $I_i^{C_0} =
C_0 \dot V_i$, which is eliminated using Kirchhoff's law
\begin{equation}
\sum_{j \nn{i}} I^s_{ij} + I^{C_0}_i + I^{R_0+n}_i = 0 ,
\end{equation}
where $I^{R_0+n}_i$ is the current through the resistor to ground,
including the noise current.
This results in
\begin{align}							\label{eq:langevin-cont}
\dot{e}_i &= \sum_{j \nn{i}} \frac{1}{2} (V_i - V_j - \dot{A}_{ij})
I^s_{ij} + V_i(-\sum_{j \nn{i}} I^s_{ij}-I^{R_0+n}_i)
\notag \\
&=-\sum_{j \nn{i}} \half (V_i + V_j) I^s_{ij} - \sum_{j \nn{i}} \half \dot{A}_{ij} I^s_{ij} - V_i I^{R_0+n}_i .
\end{align}
This is on the form of a continuity equation for the local energy,
since the sum over neighbors $j \nn i$ is the lattice analogue of a
divergence, and the source term $V_i I^{R_0+n}_i$ represents the
dissipated work done by the system on the environment. The term
involving the vector potential $A_{ij}$ will be cancelled when the
magnetization contribution is dealt with in
Sec.~\ref{sec:magnetization}.  The energy current is identified as
\begin{equation}\label{eq:JE_langevin}
I^E_{ij} = \frac{1}{2} (V_i + V_j) I^s_{ij} , 
\end{equation}
and by subtracting the transport current we obtain the heat current
\begin{equation} \label{eq:JQ_langevin}
I^Q_{ij} = \frac{1}{2} (V_i + V_j) (\Is -\Itr)
\end{equation}
for Langevin dynamics (excluding the magnetization contribution).

\subsubsection{RCSJ dynamics}

As in the Langevin case it is convenient to add a capacitance $C_0$ to
ground to the usual RCSJ model.
The energy for such a model is
\begin{equation} \label{eq:H_RCSJ}
E = -\sum_{\av{ij}} J_{ij} \cos \gamma_{ij} + \sum_{\av{ij}} \frac{1}{2}C_{ij} V_{ij}^2 + \sum_i \frac{1}{2}C_0 V_i^2 ,
\end{equation}
where $V_{ij} = \hbar \dot{\gamma}_{ij} / 2e = V_i - V_j - \dot{A}_{ij}$ is the voltage across the junction
between site $i$ and $j$. This implies a local energy of the
form
\begin{equation}
e_i = \frac{1}{2} \sum_{j \nn{i}} \left( -J_{ij} \cos \gamma_{ij} +
  \frac{1}{2} C_{ij} V_{ij}^2 \right) + \frac{1}{2} C_0 V_i^2 ,
\end{equation}
with a time derivative
\begin{equation}\label{eq:edot_rsj}
\dot{e}_i = \frac{1}{2} \sum_{j \nn{i}} \left( J_{ij}\dot{\gamma}_{ij} 
\sin \gamma_{ij} + C_{ij} V_{ij} \dot{V}_{ij} \right) + C_0 V_i \dot{V}_i .
\end{equation}
The last term is, as in the Langevin case, eliminated using current
conservation
\begin{equation}
\sum_{j \nn{i}} (I^s_{ij} + I^{R+n}_{ij} + I^C_{ij}) + I^{C_0}_i = 0 ,
\end{equation}
where the supercurrent $I^s_{ij} = I^c_{ij}\sin \gamma_{ij}$, the
current through the shunting resistor (including the noise) is $\IRn$, 
the parallel capacitance current $I^{C}_{ij} = C_{ij}\dot{V}_{ij}$, and the
current through the capacitance to ground $I^{C_0}_i = C_0\dot{V}_i$.
We get
\begin{align}
\dot{e}_i =
\sum_{j \nn{i}} (\half V_{ij} - V_i) (\Is+\Ic + \IRn) 
- \half \sum_{j \nn{i}} V_{ij} I^R_{ij} ,
\end{align} 
and by introducing the total current $\Itot = I^s_{ij} + \IRn +
I^C_{ij}$ flowing from $i$ to $j$, and rearranging
\begin{equation}\label{eq:edot_rsj2}
\dot{e}_i + \sum_{j \nn{i}} \frac{1}{2} (V_i + V_j) I^{\tot}_{ij} = -
\sum_{j \nn{i}}\half\dot{A}_{ij}I^{\tot}_{ij} - \frac{1}{2} \sum_{j
  \nn{i}} V_{ij} \IRn .
\end{equation}
These terms have interpretations similar to the Langevin case.
The second term on the left hand side is the lattice divergence of the
energy current, the first term on the right hand side will be
cancelled by the magnetization contribution, and the last one
represents the work done on the environment.
The energy current is thus
\begin{equation} \label{eq:JE_rcsj}
I^E_{ij} = \frac{1}{2} (V_i + V_j) I^{\tot}_{ij} ,
\end{equation}
and the heat current
\begin{equation} \label{eq:JQ_rsj} 
  I^Q_{ij} = \frac{1}{2} (V_i + V_j) (\Itot -\Itr)
\end{equation}
for RCSJ dynamics (again excluding the magnetization contribution).
The result is very similar to the Langevin case, except that the total
current appears instead of just the supercurrent.

\subsubsection{Magnetization contribution to the heat current}
\label{sec:magnetization}

\begin{figure}[b]
\includegraphics[width = 3cm]{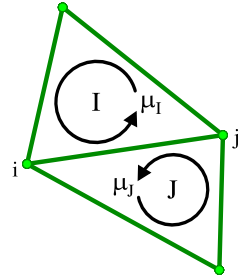}
\caption{(Color online) On each plaquette $I$ and $J$, 
defined with respect to the direct lattice points $i$ and $j$, 
there are loop currents $\mu_I$ and $\mu_J$. The difference 
between these loop currents $\mu_{ij} = \mu_I - \mu_J$ is 
the transverse part of the total current on the link from $i$ to $j$.
\label{fig:loop_current}
}
\end{figure}

The models defined in Sec.~\ref{sec:models} are formulated in the
limit where fluctuations in the vector potential are completely
suppressed, except for the uniform part $\sim \mathbf \Delta$.
Even so, the latter will, perhaps surprisingly, contribute to the heat
current.  We will derive this contribution in a more general setting
where local fluctuations in the vector potential $\bA$ are allowed.
To write down the magnetic energy, we first split the total current
flowing between sites $i$ and $j$ into transverse and longitudinal
parts
\begin{equation}							\label{eq:split}
  \Itot = 
\Itr + 
\mu_I - \mu_J + \lambda_i - \lambda_j
\end{equation}
(this is the lattice analogue of writing a vector field as the sum of
a gradient and a curl).  The variables $\mu_I$ are defined on the
plaquettes of the lattice, i.e., on the dual lattice, and are often
referred to as loop currents, see \Fig{fig:loop_current}.  The
remaining part $\lambda_i - \lambda_j$ is the loop current of the loop
$i\to j \to \text{ground} \to i$, which can be nonzero in the presence
of resistors $R_0$ and/or capacitors $C_0$ to ground.  Without loss of
generality we set $\Itr = 0$, so that the heat and energy currents
coincide.  The loop currents can be used to express the magnetic
fluxes through the corresponding loops, via the self- and mutual
inductances of the equivalent circuit diagrams
(\Fig{fig:circuit_langevin}--\ref{fig:circuit_rcsj}),
\begin{align}
& \Phi_I \equiv \sum_{J \nn I} A_{ij} = \sum_J \mathcal{L}_{IJ} \mu_J + \Phi_I^\ext
, \\
 & \Phi_{ij0} \equiv A_{ij} = \mathcal{L}_{ij} (\lambda_i - \lambda_j) ,
\end{align}
where we have chosen our gauge such that $A_{k0} = 0$, where $0$
denotes the ground, and $\Phi_I^\ext$ is the applied flux.  The sum
over $J \nn I$ is a sum over adjacent plaquettes separated by the link
$(ij)$, with $i$ and $j$ defined by \Fig{fig:loop_current}.
With this the magnetization energy is
\begin{equation}							\label{eq:EM}
  E_M = \half \sum_{I} \Phi_I \mu_I +  \half \sumij \Phi_{ij0} (\lambda_i - \lambda_j).
\end{equation}
To simplify the argument we only included the self inductances
$\mathcal{L}_{ij}$ of the
loops connecting the ground in the last term.

We may associate a local energy $e_I$ with the plaquette $I$, whose
time derivative is
\begin{align}
  \dot e_I &= \dot \Phi_I \mu_I + \half \sum_{J \nn I}  \dot \Phi_{ij0} (\lambda_i - \lambda_j)
  \notag \\
  &= \half \sum_{J \nn I} \dot A_{ij} (\mu_I - \mu_J + \lambda_i - \lambda_j)
  + \dot A_{ij} (\mu_I + \mu_J)
  \notag \\
  & = \sum_{J \nn I} \half \dot A_{ij} \Itot + \sum_{J \nn I } \half
  \dot A_{ij} (\mu_I + \mu_J) .
\end{align}
The first term in the last line cancels exactly the corresponding
contribution in Eqs.~\eqref{eq:langevin-cont} and
\eqref{eq:edot_rsj2}.  The summation in the last term represents the
energy flowing into the plaquette $I$ from the adjacent plaquettes,
and we can therefore identify the magnetization contribution to the
energy current and hence the heat current
\begin{equation}							\label{eq:IQM}
  I_{IJ}^Q =- \half \dot A_{ij} (\mu_I + \mu_J) .
\end{equation}
It is not hard to see that this result holds also in presence of an
electric transport current, if the $\mu_I$:s are defined as in
\Eq{eq:split}.
Equation \eqref{eq:IQM} is the discrete analogue of the last term of
\Eq{eq:AxM}.

\subsubsection{Full heat current}

Averaging \Eq{eq:JQ_langevin} or \eqref{eq:JQ_rsj} over the system and
adding the magnetization contribution \Eq{eq:IQM} gives the full
average heat current density in the $x$-direction
\begin{align}								\label{eq:JQ_tot}
\bar{J}^Q_{x} =& \frac{1}{\Omega}
  \sum_{\av{ij}} x_{ji} \half  (V_i + V_j)(\Itot - \Itr)
\notag \\
-& \frac{1}{\Omega} \sum_{\av{IJ}} \left[ \left( x_{ij}^c - x_I \right) \mu_I 
   - \left( x_{ij}^c - x_J \right) \mu_J \right] \dot A_{ij} ,
\end{align}
where $x_{ji} = x_j - x_i$ is the $x$-component of the difference vector
from site $i$ to $j$, $x_{ij}^c = (x_i + x_j)/2$, and $x_I$ the
plaquette centers (the vertices of the Voronoi graph).
In the limit where $\dot \bA$ is spatially uniform the last term
simplifies to $-\dot A_y M_z$, where $M_z = \sum_I \mu_I \Omega_I /
\Omega$ is the average magnetization density [$\Omega_I$ is the area
of the (Voronoi) cell $I$], in agreement with \Eq{eq:AxM}.  A more
practically useful form is obtained below in
Sec.~\ref{sec:Kubo-magnetization} by expressing the loop currents
$\mu_I$ in terms of the total currents $\Itot$, which results in the
expression \Eq{eq:heat-current}.

\subsection{Heat current from Kubo formula}
\label{sec:Kubo}

It is useful to see how the heat current enters the linear response
arising from an applied temperature gradient via a Kubo formula.  We
will do this starting from the stochastic equations of motion of
either Langevin [\Eq{eq:langevin1}] or RCSJ [\Eq{eq:RCSJ1}]
dynamics. The temperature enters these equations only via the strength
of the Gaussian noise correlation function, which now gets a spatial
dependence $T(x) = T_0 - T' x$.~\footnote{In reality the parameters of
  the model, e.g., $\Ic$, are going to be temperature dependent and
  therefore also spatially dependent in a temperature
  gradient. However, this dependence only changes the equilibrium
  distribution of the model and does not give rise to a thermodynamic
  force. We can therefore ignore this temperature dependence in the
  derivation of the heat current.}  $T'$ is considered a weak
perturbation, and we are interested in calculating the response of
some dynamical variable $A$ to such a perturbation.
We assume open boundary conditions in the $x$-direction, i.e., in the
direction of the applied temperature gradient, while in the transverse
$y$-direction they can be arbitrary.
With no net electric current flowing through the sample
the heat current is equal to the energy current.

For concreteness we consider the response of the electric field
$E_y$ perpendicular to the temperature gradient and the magnetic
field, which gives the Nernst signal $e_N = \delta\av{E_y}/\delta T'$.
Our goal is to express this linear response using a Kubo formula
\begin{equation}				\label{eq:Kubo}
  e_N = \frac{\Omega}{T^2} \int_0^\infty \av{\bar E_y(t) \bar J_x^Q(0)} dt ,
\end{equation}
where $\bar J_x^Q$ then can be identified with the heat current (the
overbar denotes a spatial average), and
$\Omega=L_x L_y$ is the system size.

To this end is convenient to reformulate the problem using functional
integrals~\cite{Janssen1976} and write ensemble averages
as
\begin{equation}				\label{eq:func-int}
  \av{A(t)} = \frac{1}{Z} \int A(\theta(t)) e^{-S[\theta]}
  J[\theta] [d\theta] ,
\end{equation}
where $e^{-S}$ is the statistical weight of a given realization of
the stochastic process $\theta(t)$. The Jacobian $J[\theta]$ in
\Eq{eq:func-int} comes from the transformation from the Gaussian
noise $\zeta$ to $\theta$ \cite{Janssen1976}, but since it plays no
role in the following it will be dropped henceforth.
The linear response of $A$ to a temperature gradient can now be
expressed (assuming time translation invariance) as
\begin{equation}					\label{eq:KuboQ0}
  R(t-t') = \frac{\delta \av{A(t)}}{\delta T'(t')} = \av{A(t) Q(t')} ,
\end{equation}
where
\begin{equation}
  Q(t') = - \left. \frac{\delta S[\theta]}{\delta T'(t')} \right|_{T'=0} .
\end{equation}
We are interested in the response to a static perturbation in a
stationary state given by
\begin{equation}					\label{eq:KuboQ}
  R = \int_{-\infty}^\infty \av{A(t)Q(0)} dt .
\end{equation}
Despite the similarity with \Eq{eq:Kubo}, we cannot immediately
identify $2T^2 Q(t)/\Omega$ with $J^Q(t)$ at this stage, because of their
different symmetry properties, an issue we now digress into.

\subsubsection{Symmetry relations of transport coefficients and
  correlation functions}

Before continuing the derivation of the heat current from the Kubo
formula we discuss some general properties of linear response in
classical statistical systems.  It is well known that time reversal
symmetry implies symmetry relations among the transport coefficients
and among correlation functions.  In the presence of a magnetic field
$H$ the time reversal operation should reverse also that.
Consider now the correlation function $C_{E_yQ}(t,H) = \av{E_y(t) Q(0)}$
entering \Eq{eq:KuboQ0}, which in general also depends on the magnetic
field $H$.  The electric field $E_y$ is even under time reversal,
while $Q$ is neither even nor odd.
Instead we may split $Q = Q_e + Q_o$ into even and odd parts.
By parity symmetry the correlation functions $C_{E_y Q_{e,o}}$ are odd
in $H$, so that
\begin{align}
 \!\! C_{E_y Q_e}(t,H) &= C_{E_y Q_e}(-t,-H) =-C_{E_y Q_e}(-t,H), \\
 \!\! C_{E_y Q_o}(t,H) &= -C_{E_y Q_o}(-t,-H) =C_{E_y Q_o}(-t,H) .
\end{align}
This leads to the following symmetry of the response
\begin{align}
  R(t,H) & = C_{E_yQ_e}(t,H) + C_{E_yQ_o}(t,H) \nonumber \\
 &=  -C_{E_yQ_e}(-t,H) + C_{E_yQ_o}(-t,H) .
\end{align}
For $t<0$ causality implies that $R(t,H) = 0$, hence
$C_{E_yQ_o}(-t,H) = C_{E_yQ_e}(-t,H)$, so that the response can be written
solely in terms of the odd part of $Q$ as
\begin{equation}
 R(t,H) = 2\Theta(t) C_{E_yQ_o}(t,H),
\end{equation}
where $\Theta(t)$ is the Heaviside step function.
For the Nernst signal we then get
\begin{equation}
  e_N = \int_0^\infty 2 C_{E_y Q_o}(t) dt
  = \int_{-\infty}^\infty \av{E_y(t) Q_o(0)} dt .
\end{equation}
Thus, in order to evaluate the response it is enough to consider the
odd part of $Q(t)$.

Let us also make the following observation:
At finite times the response will contain transients, which we will not
be interested in and which do not contribute to the stationary
response.
Indeed, any contribution to $Q_o(t)$ of the form of a total time
derivative $df/dt$ will contribute a term
\begin{equation}
  \int_{-\infty}^{\infty} \langle E_y(0) \frac{df(t)}{dt} \rangle dt =
  \av{E_y(0)f(\infty)} - \av{E_y(0)f(-\infty)}
\end{equation}
to $e_N$, which vanishes provided $f(t)$ is stationary in equilibrium.
We will now treat the Langevin and RCSJ case separately.

\subsubsection{Langevin dynamics}

For the case of Langevin dynamics the dynamical action $S$ obtains
from substituting $\eta$ using \Eq{eq:langevin1} in the Gaussian noise
probability distribution
\begin{equation}
 P[\eta] \propto \exp\big(-\int\sum_i \frac{\hbar}{4 \gamma T_i}
   \eta_i^2(t) dt \big)
\end{equation}
resulting in
\begin{equation}				\label{eq:S-Langevin}
  S_\text{Lang.} = \int  \sum_i 
   \frac{\hbar}{4 \gamma T_i} \left( \gamma \dot\theta_i - F_i \right)^2 dt,
\end{equation}
where
\begin{equation}
 F_i = - \frac{1}{\hbar} \frac{\partial H}{\partial \theta_i} = - \frac{1}{2e}
 \sum_{j \nn i} \Is ,
\end{equation}
and correspondingly
\begin{equation}
 Q(t) = - \sum_i \frac{\hbar x_i}{4 \gamma T^2}
  \left( \gamma \dot\theta_i(t) - F_i(t) \right)^2 .
\end{equation}
Since $\dot \theta_i = 2eV_i/\hbar$ is even and $\Is$ is odd under
time reversal~\footnote{$\theta$ changes sign under time reversal},
the odd part of $Q$ is
\begin{align}
  Q_o(t) &= \frac{1}{2 T^2} \sum_i x_i \hbar \dot\theta_i F_i \nonumber \\
	&= - \frac{1}{2 T^2} \frac{\hbar}{2e}\sum_i x_i \dot\theta_i
        \sum_{j \nn i} \Is  \nonumber \\
	&= - \frac{1}{2 T^2} \frac{\hbar}{2e} \sumij (x_i \dot\theta_i -
        x_j \dot\theta_j ) \Is .
\end{align}
Putting $x_{ij}^c = (x_i + x_j)/2$ we may rewrite this as
\begin{align}
2 T^2 Q_o(t) = & - \sumij x_{ij}^c (V_i - V_j - \dot A_{ij}) \Is
\notag \\
& + \sumij \left( \half x_{ji} (V_i + V_j) - x_{ij}^c \dot A_{ij}
\right) \Is .
\end{align}
The first term is a total time derivative of $x_{ij}^c$ times a local
energy $e_{ij}$
\begin{equation}
  \frac{d}{dt} \sumij x_{ij}^c e_{ij} = - \sumij x_{ij}^c \frac{d}{dt}
  J_{ij} \cos \gamma_{ij} ,
\end{equation}
and will therefore not contribute to static response functions.
The remaining part can be identified with the heat current in the
Langevin model,
\begin{equation}								\label{eq:JQ1}
\bar J^Q_x = \frac{1}{\Omega} \sumij \left( \half x_{ji} (V_i + V_j) -
  x_{ij}^c \dot A_{ij} \right) \Is .
\end{equation}
This form of the heat current is directly formulated using the
currents and potentials, and therefore simpler to use than \Eq{eq:JQ_tot}.
To show the equivalence of Eqs.~\eqref{eq:JQ1} and \eqref{eq:JQ_tot}
it is necessary to go through some further steps.
Before doing that, however, we consider the RCSJ case.

\subsubsection{RCSJ dynamics}

It is convenient to reformulate the equations of motions for RCSJ
dynamics \Eq{eq:RCSJ1} as
\begin{align}							\label{eq:RCSJ-2}
& \Itot = \Ic + \Is + \IR + \In = \Itr + \lambda_i - \lambda_j + \mu_{ij}, \\
									\label{eq:RCSJ-3}
& \sum_{j \nn i} \lambda_i - \lambda_j = C_0 \dot V_i ,
\end{align}
where $\lambda_i - \lambda_j$ is the longitudinal part of the total
current flowing through the links of the lattice, and $\mu_{ij} =
\mu_I - \mu_J$ is the transverse part, with the $\mu_I$ defined on the
dual lattice sites, i.e., on the plaquettes, adjacent to the bond $ij$
as in \Fig{fig:loop_current}. The $\mu_I$:s are often referred to as
loop currents.  Neither of these contribute to the transport current
$\Itr$.  As in the Langevin case we set $\Itr = 0$, whereby the heat
current equals the energy current.

The dynamical action corresponding to
Eqs.~(\ref{eq:RCSJ-2}--\ref{eq:RCSJ-3}) can be expressed as
\begin{equation} \label{eq:S-RCSJ-1}
  S'_\text{RCSJ} = \int \sum_{\av{ij}} \Big\{ \frac{R_{ij}}{4T_{ij}} (\In)^2
+ i \eta_{ij} (\Itot - \lambda_i + \lambda_j - \mu_{ij} ) \Big\} dt .
\end{equation}
The first term represents the Gaussian distribution of the white noise
current $I_{ij}^n$, and $\eta_{ij}$ is a Lagrange
multiplier to enforce the constraints \Eq{eq:RCSJ-2}.
In a temperature gradient the local temperatures are position
dependent, $T_{ij}=T - T' x_{ij}^c$.

The functional integration is over the variables $\theta$, $\In$,
$\eta_{ij}$, $\lambda_i$, and $\mu_I$. Integrating over $\In$ and
$\eta_{ij}$ we get
\begin{equation} \label{eq:S-RCSJ}
 S_\text{RCSJ} = \int \sum_{\av{ij}} \frac{R_{ij}}{4T_{ij}} ( \Ic + \Is + \IR
- \lambda_i + \lambda_j - \mu_{ij} )^2 dt ,
\end{equation} 
and
\begin{equation}
  Q(t) = - \frac{1}{4T^2} \sum_{\av{ij}} x_{ij}^c R_{ij} 
  (\Ic + \Is + \IR - \lambda_i + \lambda_j - \mu_{ij})^2.
\end{equation}
Again only the part which is odd under time reversal contributes to the
heat current,
\begin{equation}
 2 T^2 Q_o = - \sumij x_{ij}^c V_{ij} \left( \Ic + \Is -\lambda_i +
   \lambda_j - \mu_{ij} \right),
\end{equation} 
since $\IR = V_{ij}/R_{ij}$ is even, while the other currents are odd.
In this expression we may identify a contribution 
\begin{equation}
  \sumij x_{ij}^c V_{ij} (\Ic + \Is) = \frac{d}{dt} \sumij x_{ij}^c e_{ij} ,
\end{equation}
where
\begin{equation}
e_{ij} = \half C_{ij} V_{ij}^2 - J_{ij} \cos(\theta_i - \theta_j -
\frac{2\pi}{\Phi_0} A_{ij})
\end{equation}
is the local energy defined on the links of the lattice.
Being a total time derivative this does not contribute in a stationary
state.
The remaining part can be rearranged into
\begin{align*}
 & \sumij x_{ij}^c V_{ij} \left( \lambda_i - \lambda_j + \mu_{ij} \right) \\ 
& = \sum_i x_i V_i  \sum_{j \nn i} \left( \lambda_i - \lambda_j +
  \mu_{ij} \right)  \\
&+  \sumij \left[ x_{ji} \half (V_i + V_j) - x_{ij}^c \dot
 A_{ij}\right] \left( \lambda_i - \lambda_j + \mu_{ij} \right) .
\end{align*}
The second line equals $\sum_i x_i C_0 V_i \dot V_i = \frac{d}{dt} \sum_i
x_i \half C_0 V_i^2$, again a total time derivative which does not
contribute.  Finally, $\lambda_i - \lambda_j + \mu_{ij} = \Itot$, so
that the heat current for the RCSJ model becomes
\begin{equation}								\label{eq:JQ2}
\bar J^Q_x = \frac{1}{\Omega} \sumij \left( x_{ji} \half (V_i + V_j) - x_{ij}^c \dot A_{ij}
\right) \Itot .
\end{equation}

\subsubsection{Magnetization contribution}
\label{sec:Kubo-magnetization}

Equations \eqref{eq:JQ_tot} and \eqref{eq:JQ1}, \eqref{eq:JQ2}
apparently differ. We will now show the equivalence of these
formulations.  We split the current into transverse and longitudinal
parts as in \Eq{eq:split}, with the $\mu_I$ defined on the dual
lattice, whose sites are denoted by $\br_I = (x_I, y_I)$.
With this it is possible to rewrite the last term in
Eqs.~\eqref{eq:JQ1},~\eqref{eq:JQ2}, as
\begin{align*}
- \sumij x_{ij}^c \dot A_{ij} \Itot =
- \sumij x_{ij}^c \dot A_{ij} (\lambda_i - \lambda_j + \mu_{ij}) 
\\
=  - \sumij x_{ij}^c \dot A_{ij} (\lambda_i - \lambda_j)
- \sum_I x_I \mu_I \sum_{J \nn I} \dot A_{ij}
\\
- \sumIJ \left[ \left( x_{ij}^c - x_I \right) \mu_I 
  - \left( x_{ij}^c - x_J \right) \mu_J \right] \dot A_{ij} .
\end{align*}
The first term is a total time derivative of
$\sumij x_{ij}^c e_{ij}^M$, where $e_{ij}^M = \half A_{ij}
(\lambda_i - \lambda_j)$ is the local magnetization energy of
the loops to ground [cf. \Eq{eq:EM}].
The second term is a total time derivative of $\sum_I x_I e_I^M$,
involving the magnetization energy $e_I ^M = \half \mu_I \Phi_I$ of
the loops of the lattice.  The remaining part corresponds exactly to
the last term of \Eq{eq:JQ_tot}.  Note that for the models discussed
initially $\dot \bA = (\Phi_0/2\pi) \dot {\mathbf{\Delta}}$, i.e.,
only spatially uniform fluctuations are included in the vector
potential.  Then $\dot e_{ij}^M = \dot e_I = 0$ exactly, and
Eqs.~\eqref{eq:JQ_tot} and \eqref{eq:JQ1}, \eqref{eq:JQ2} become
identical (for open boundary conditions).  In the more general case
where local fluctuations are allowed they differ only by a total time
derivative, which does not contribute to the transport coefficients.

\subsection{Additional remarks}

The two derivations of the heat current given in
Sec.~\ref{sec:conserv} and \ref{sec:Kubo} above agree. The latter one
shows, in addition, the validity of the standard Kubo formula
\Eq{eq:nernst_kubo} for calculating the Nernst response of the
transverse electric field to an applied temperature gradient (in
absence of an electric transport current).  The resulting expression
should also give, via an Onsager relation, the response of the heat
current to an applied transverse electric current.  This holds
provided the electric transport current $\Itr$ is subtracted from the
current in Eqs.~\eqref{eq:JQ1}, \eqref{eq:JQ2}, showing that
\Eq{eq:heat-current} is indeed the correct form.

In both the Langevin and the RCSJ case the heat current is given by
similar expressions, but in the RCSJ case the current
$I_{ij}^\text{tot}$ includes also capacitative, resistive, and noise
currents in addition to the supercurrent.

As mentioned, Eq.~\eqref{eq:heat-current}, being directly formulated
in the currents and phases, have a clear advantage over
\eqref{eq:JQ_tot}.  They are equivalent for systems with open boundary
conditions along the temperature gradient.  In simulations periodic
boundary conditions are convenient to use in order to eliminate
surface effects. This, however, makes things more subtle as then the
magnetization is not uniquely determined by the currents: Adding a
constant to every $\mu_I$ does not change $\Itot$ in \Eq{eq:split}.
One logical possibility seems to be to impose an extra condition on
the average magnetization, e.g., define it to be zero at any moment,
or $\sum_I \mu_I = 0$, so that no magnetization contribution should be
added in this case.  Another option is to fix one particular $\mu_I$,
and in effect use \Eq{eq:heat-current} also for periodic boundary
conditions.  We opt for this latter condition, since it stays closer
to the experimental open system situation, while getting rid of
surface effects.  This choice can be further justified by comparing
analytic results for open and periodic boundary conditions obtained in
a spin wave approximation, to be discussed next.

\section{Heat conductivity at low temperature and zero magnetic field}

\label{sec:linearized}

At low enough temperatures and zero magnetic field the fluctuations
will be small so that it is sufficient to consider linearized versions
of the models introduced earlier.  The only nonlinear circuit element
is the Josephson junction, and by linearizing the Josephson relation
$\Is = I_{ij}^c \sin \gamma_{ij} \approx I_{ij}^c \gamma_{ij}$, the
models are reduced to a network of capacitors, resistors and
inductors, with effective inductances $\mathcal L_{ij} = (\hbar/2e
I_{ij}^c)$.  In this spin wave approximation vortices are absent,
hence there will be no Nernst effect, but the thermal conductivity
will still be nonzero.  For the Langevin case the resulting model can
be mapped to one of heat conduction by phonons in a harmonic crystal
coupled to local heat baths, which has an analytic
solution~\cite{Bonetto2004}.  For a 2D infinite square lattice the
spin wave heat conductivity is independent of temperature and given by
\begin{align}						\label{eq:kappa-langevin-linearized}
  \kappa_\text{sw} &= \frac{k_B I_c}{2e \gamma} \int_0^1 \int_0^1
  \frac{\sin^2 (\pi x)}{4 \sin^2(\frac{\pi x}{2}) + 4 \sin^2(\frac{\pi y}{2})}dxdy
\nonumber \\
  &=  (\frac{1}{2} - \frac{1}{\pi})  \frac{k_B I_c}{2e \gamma}  \approx 0.1817  \frac{k_B I_c}{2e \gamma} .
\end{align}

We now turn to the heat conductivity of the linearized RCSJ model on a
2D square array of size $\Omega=L_x L_y$ and unit lattice constant.
In this case, the heat current [\Eq{eq:JQ2}] includes the total
current $\Itot = \Ic + \IR + \Is + \In$, which is purely transverse
due to Eqs.~\eqref{eq:RCSJ1} and \eqref{eq:RCSJ2}.  In this sum, $\Ic$
and $\IR$ have no transverse component, so that $\Itot =
I^{s\perp}_{ij} + I^{n\perp}_{ij}$.  The transverse part of the
supercurrent $I^{s\perp}_{ij}$ is entirely due to vortices and
vanishes in the spin wave approximation, so that the total current
only consists of the transverse component of the noise current, $\Itot
= I^{n\perp}_{ij}$.  More explicitly, this result can be derived from
the equations of motion.  On a square lattice it is convenient to
label the links by the coordinate $\br$ and direction $\mu=x,y$, and
introduce forward and backward difference operators $\df_\mu f(\br) =
\db_\mu f(\br+\hat{\bm{\mu}}) = f(\br+\hat{\bm{\mu}}) -f(\br)$.
Introducing rescaled variables $\phi_i = \hbar\theta_i/ 2e$ and
$\tilde \Delta = \hbar \Delta / 2e$ the equations of motion
\eqref{eq:RCSJ1}, \eqref{eq:RCSJ2} are in this limit
\begin{align}								\label{eq:RCSJ_eomlin1}
\sum_\mu  \db_\mu\df_\mu  \left( C\ddot \phi_\br + \frac{1}{R} \dot \phi_\br +
  \frac{1}{\mathcal L} \phi_\br \right)  &= \sum_\mu \db_\mu I^n_{\br \mu} ,
\\
 C \ddot{\tilde\Delta}_\mu + \frac{1}{R}\dot{\tilde\Delta}_\mu +
\frac{1}{\mathcal L} \tilde\Delta_\mu &= \frac{1}{\Omega} \sum_{\br} I^n_{\br \mu} \label{eq:RCSJ_eomlin2} .
\end{align}
Multiplying \Eq{eq:RCSJ_eomlin1} with the lattice Green's function $G$
(solving $-\db\df G_{\br\br'} = \delta_{\br,\br'}$) gives
\begin{equation}
  \label{eq:RCSJ_eomlin1_alt}
  C \ddot{\phi}_\br +
  \frac{1}{R}\dot{\phi}_\br + \frac{1}{\mathcal L} \phi_\br = -\sum_{\br'\nu}
  G_{\br\br'} \db_\nu I^n_{\br'\nu} .
\end{equation}
Using \eqref{eq:RCSJ_eomlin1_alt} and \eqref{eq:RCSJ_eomlin2}, the
total current on a square lattice is
\begin{align} \label{eq:RCSJ_lincurr1}
I_{\br\mu}^{\text{tot}} & = I_{\br\mu}^{C} +
I_{\br\mu}^{R} + I_{\br\mu}^s + I_{\br\mu}^n
\notag \\
& = I_{\br\mu}^{n} + \sum_{\br'\nu}  \df_\mu G_{\br\br'} \db_\nu I^n_{\br'\nu}  -
\frac{1}{\Omega} \sum_{\br'} I^n_{\br'\mu} .
\end{align}
We can here identify the longitudinal part of the noise current
$I^{n \parallel}_{\br\mu} = -\sum_{\br'\nu} \df_\mu G_{\br\br'}
\db_\nu I^n_{\br'\nu}$, and the average ($\bk = 0$ component)
$I_{\mu}^{n0} = \frac{1}{\Omega} \sum_{\br} I^n_{\br\mu}$.  The total
current is thus just the transverse part of the noise current
\begin{equation} \label{eq:RCSJ_lincurr}
 I_{\br\mu}^{\text{tot}} = I^n_{\br\mu} - I^{n \parallel}_{\br\mu} -
 I_{\mu}^{n0} = I^{n \perp}_{\br\mu} .
\end{equation}
The heat current in the $x$-direction can then be written as
\begin{align}
 \bar J^Q_x &= \frac{1}{\Omega} \sum_{\br\nu} \chi_{\br\nu} I^{n\perp}_{\br\nu} ,
 \\
 \chi_{\br\nu} &= \half(\dot \phi_\br + \dot
 \phi_{\br+\hat{\bx}})\delta_{\nu x} - 
 (x + \half \delta_{\nu x})\dot{\tilde{\Delta}}_\nu .
\end{align}
This result is intriguing because the dynamics of the system is
completely independent of $I^{n\perp}_{\br\mu}$, as seen from the
equations of motion.  Correspondingly, the correlation function which
enters the Kubo formula \Eq{eq:kappa_kubo} factorizes:
\begin{align}
  & \av{\bar J^Q_x(t)\bar J^Q_x(0)}
  = \frac{1}{\Omega^2}
  \sum_{\br\mu, \br'\nu}  
\av{\chi_{\br\mu}(t) I^{n\perp}_{\br\mu}(t)
    \chi_{\br'\nu}(0) I^{n\perp}_{\br' \nu}(0)}
  \notag \\								\label{eq:JQJQ}
  &\qquad = \frac{1}{\Omega^2}
  \sum_{\br\mu, \br'\nu}
\av{\chi_{\br\mu}(t) \chi_{\br'\nu}(0)} \av{I^{n\perp}_{\br\mu}(t) I^{n\perp}_{\br' \nu}(0)}
.
\end{align}
Furthermore, the transverse noise current correlation function
\begin{multline}
  \label{eq:InperpInperp}
  \big< I^{n \perp}_{\br \mu}(t) I^{n \perp}_{\br' \nu}(0) \big>  \\ =
  \frac{2 k_B T}{ R} \delta(t)
  \frac{1}{\Omega} \sum_{\bk \neq 0}
  \left( \delta_{\mu \nu} - \frac{\ka_\mu \ka_\nu}{\ka^2}
  \right) e^{i\bk \cdot(\br -\br')} ,
\end{multline}
where $\ka_\mu = 2\sin(k_\mu/2)$, $\ka^2 = \sum_\mu
\ka_\mu^2$, $k_\mu = 2\pi n_\mu/L_\mu$, and $n_\mu = 0, \ldots ,
L_\mu-1$.  Since this is proportional to $\delta(t)$, the correlation
function $\av{\chi(t)\chi(0)}$ has to be evaluated at $t = 0$ when
inserted into the Kubo formula, i.e., it is given by an
\emph{equilibrium} correlation function.

In the more general case the total current contains also the
transverse part of the supercurrent, which is determined by the
vortices.
Clearly the vortex contribution appears on top of the spin wave
background calculated in this section.
We may evaluate the correlation function
$\av{\chi_{\br\mu}\chi_{\br'\nu}}$ for the full model, including a
capacitance $C_0$ to ground and cosine interaction.
On a square lattice the RCSJ Hamiltonian \Eq{eq:H_RCSJ} is
\begin{equation} \label{eq:H_RCSJ_r}
H = \sum_\br \frac{1}{2}C_0 V_\br^2 + 
\sum_{\br,\mu} \frac{1}{2} C
(V_\br - V_{\br + \hat{\mu}} - \dot{\tilde{\Delta}}_\mu)^2
+ f(\phi, \tilde{\bm{\Delta}}) ,
\end{equation}
where $V_\br = \dot \phi_\br$ and $f(\phi, \tilde{\bm{\Delta}})$ is
the ``potential energy'' involving the cosine interaction.
Switching to Fourier space
\begin{equation} \label{eq:H_RCSJ_k}
H = \frac{1}{\Omega} \sum_\bk \frac{1}{2}(C_0 + C \ka^2)V_{-\bk}V_{\bk} + 
\frac{1}{2} C \Omega (\dot{\tilde\Delta}_x^2 + \dot{\tilde\Delta}_y^2)
+ f(\phi, \tilde{\bm{\Delta}}) .
\end{equation}
From here it is easy to calculate the required averages, since the
partition function factorizes at the classical level and the averages
are just Gaussian integrals.  We get
\begin{equation} \label{eq:VVcorr}
 \av{V_{-\bk}V_{\bk}} = \frac{k_B T}{C_0 + C\ka^2}, 
\qquad
 \big< \dot{\Delta}_\mu \dot{\Delta}_\nu \big> = \frac{k_B T}{C \Omega}
 \delta_{\mu \nu} ,
\end{equation}
independent of $f(\phi,\tilde{\bm{\Delta}})$, so that
\begin{align}									\label{eq:chichi}
  \av{\chi_{\br x} \chi_{\br' x}} &= \frac{1}{\Omega} \sum_\bk e^{i \bk
   \cdot (\br - \br')} \frac{1}{4} \left| 1 + e^{i k_x} \right|^2
  \av{V_\bk V_{-\bk}}
\notag \\
  &+ \Big( x + \half \Big) \Big( x' + \half \Big)
 \av{\dot{\tilde\Delta}_x \dot{\tilde\Delta}_x} ,
\\[3mm]
 \av{\chi_{\br y} \chi_{\br' y}} &=   x x' \av{\dot{\tilde\Delta}_y \dot{\tilde\Delta}_y} ,
\\ \label{eq:chichi-last}
 \av{\chi_{\br x} \chi_{\br' y}} &= \av{\chi_{\br y} \chi_{\br' x}} = 0 .
\end{align}
Performing the sum over $\br, \br'$ in \eqref{eq:JQJQ} and using
\eqref{eq:InperpInperp}, \eqref{eq:VVcorr}--\eqref{eq:chichi-last} 
the heat conductivity becomes
\begin{align}								\label{eq:kappa}
 \kappa_\text{sw} &= \kappa' + \kappa'',
  \\										\label{eq:kappa'}
 \kappa' &= \frac{k_B}{RC} \frac{1}{\Omega} \sum_{\bk \neq 0} \frac{\left( 1 + \frac{1}{4}
      \ka_x^2\right) \left( 1 - \frac{\ka_x^2}{\ka^2} \right)}{C_0/C + \ka^2} ,
  \\										\label{eq:kappa''}
  \kappa'' &= 
\frac{k_B}{RC} \frac{L_x^2 -1}{12 \Omega} .
\end{align}
The second term $\kappa''$ originates from
$\langle\dot{\tilde\Delta}_y\dot{\tilde\Delta}_y\rangle$ and
corresponds to the magnetization contribution.

The resulting heat conductivity \Eq{eq:kappa} has some notable
properties.  Firstly, it is proportional to $1/RC$ so that it is well
defined only for finite $C$.  In this respect the RSJ model, without
shunting capacitors, i.e., with $C \to 0$ is pathological.  Secondly,
when $C_0 = 0$ the sum over $\bk$ in \eqref{eq:kappa'} is
logarithmically divergent in the infinite system $L \to \infty$ (in
2D).  For finite large $L$ the heat conductivity has a logarithmic
size dependence
\begin{equation}								\label{eq:kappa_large_L}
  \kappa_\text{sw} \sim \frac{1}{4\pi RC} \ln \frac{L}{a} ,
\end{equation}
where $a$ is the lattice spacing.  $\kappa$ is thus not a bulk
property of the RCSJ model in 2D.  It is interesting to note that
several other low dimensional models of heat conduction display an
anomalous size dependence, often tied to momentum
conservation~\cite{Lepri2003,Dhar2008}.  In the present case the
diverging behavior is most likely due to the long range Coulomb
interaction.  A finite $C_0$ makes the charge-charge interaction
exponentially small on distances larger than the screening length
$\lambda = a \sqrt{C/C_0}$, and also yields a system size independent
$\kappa_\text{sw}$ when $L \gg \lambda$.
The calculation above was done for periodic boundary conditions. We
have repeated it for open boundary conditions in the
$x$-direction. The difference is very small and tends to zero as
$1/L_x$ when $L_x$ increases.

As discussed above, the vortex contribution appears on top of the
temperature independent spin wave background just calculated.
Computationally it is often convenient to project out the spin wave
contribution by excluding the transverse noise current $I^{n\perp}$ in
\Eq{eq:JQ_tot} before the averaging in \Eq{eq:kappa_kubo}.

\section{Simulation methods}					\label{sec:methods}

The equations of motion for Langevin dynamics [\Eq{eq:langevin1} and
\Eq{eq:langevin2}] are solved numerically using a simple forward Euler
discretization with a time step of $\Delta t = 0.02$. The RCSJ
dynamical equations [\Eq{eq:RCSJ1} and \Eq{eq:RCSJ2}] have a more
complicated structure and are also second order in time, which makes
the solution numerically more intensive. To solve these we use a
leap-frog type discretization scheme, with time step $\Delta t =
0.04$. For a system of $N$ grains one then generally has to solve a
system of $N+1$ coupled equations in each time step ($N-1$ for the
phases $\{ \theta_i \}$ and 2 for the twists $\bm{\Delta}$).  Note,
however, that the equation system is sparse, so an effective way to
solve them is to employ an LU factorization algorithm, since the
complexity of such a scheme goes as the number of nonzero entries,
which are of the order of $N$ here. This is far better than the direct
method~\cite{MonTeitel,ChungLeeStroud} of multiplying with the lattice
Green's function, which is here a dense matrix with $\sim N^2$
nonzero entries.  In both the Langevin and the RCSJ case the sampling
is performed during $4 \cdot 10^5$ units of time, after an
equilibration time of 10\% of this.

We have simulated systems of sizes up to $120\times120$, but except
the case of $\kappa$ for RCSJ dynamics, finite size effects are
unimportant for systems larger than $L \gtrsim 20$, and thus only
systems of size $20\times20$ are considered.

The transport coefficients are calculated from the Kubo formulas
Eqs.~\eqref{eq:nernst_kubo}-\eqref{eq:rho_kubo},
where the upper limit in the time integrals is replaced by a large
enough time ($\gtrsim$ the correlation time), such that the 
cumulative value of the integral has stabilized its value.
The Nernst signal $e_N$ obtained from the Kubo formula in
\Eq{eq:nernst_kubo} is also compared with the value of $e_N$
calculated in two other ways. The first is simply to apply a small
temperature gradient in the $x$-direction and measure the resulting
electric field $E_y$, and the second is through the Onsager relation
$e_N = \tilde\beta_{xy} / T$, where $\tilde\beta$ is the response of
the heat current in the $x$-direction to an applied electric current
in the $y$-direction, as defined in \Eq{eq:thermoelectric2}. 
This gives an important consistency check of our calculations.  A
similar check is performed for the heat conductivity $\kappa$.  The
Kubo formula turns out to be the most efficient way to calculate the
response, since one does not have to worry about nonlinear effects.

\subsection{Units}
In the simulations we redefine time and temperature to make them
dimensionless (we also measure length in units of a lattice
constant $a$). For both Langevin and RCSJ dynamics, temperature is
rescaled according to
\begin{equation}
  T \to T \frac{2e k_B}{\hbar I^c}.
\end{equation}
The dimensionless time is obtained from the transformation 
\begin{equation}
 t \to t \frac{I^c}{2e \gamma},
\end{equation}
for Langevin dynamics. In the RCSJ case the rescaling is
\begin{equation}
  t \to t \frac{2e R I^c}{\hbar} .
\end{equation}
(In the case where the junction parameters vary from link to link
$I^c$, $R$ and $C$ denote a characteristic magnitude.)
From this follows that for Langevin dynamics, the Nernst signal $e_N$
is given in units of $k_B / 2e\gamma$, the thermal conductivity
$\kappa$ in units of $k_B I^c / 2e\gamma$, and the resistivity $\rho$
in units of $\hbar / (2e)^2 \gamma$. For RCSJ dynamics the Nernst
signal $e_N$ is measured in units of $ 2e k_B R / \hbar$, the thermal
conductivity $\kappa$ in units of $2e k_B R I^c / \hbar$, and the
resistivity $\rho$ in units of the shunt resistance $R$.
The dimensionless parameter $Q^2 = 2e R^2 I^c C / \hbar$ (the ratio of
the two times scales $RC$ and $\hbar/ 2eR I^c$) controls the
damping. For $Q \gg 1$ the system is underdamped and for $Q \ll 1$ it is
overdamped.

\subsection{Time discretization of the heat current} \label{sec:discheat}

It is crucial to use a symmetric time discretization of the heat
current, \Eq{eq:heat-current}, in the numerics.  For Langevin
dynamics, while it is sufficient to use a forward Euler discretization
for the integration of the equations of motion, the voltage $V_i =
\hbar \dot \theta_i/2e$ appearing in \Eq{eq:heat-current} has to be
approximated by a centered difference $\dot \theta(t) \approx
\{\theta(t+\Delta t) - \theta(t-\Delta t)\}/2\Delta t$.  In the RCSJ
case the total electric current is (after rescaling time)
\begin{equation} \label{eq:Itott}
  I_{ij}^{\text{tot}}(t) = I_{ij}^c \left( \sin \gamma_{ij}(t) +
    \dot{\gamma}_{ij}(t) + Q^2 \ddot{\gamma}_{ij}(t) \right) + \In(t) .
\end{equation}
In the symmetric leap-frog scheme we use, $\theta$ is defined on
integer time steps $t = n \Delta t$, while the first order time
derivative $\dot \theta$ is defined only on half-integer time steps $t
= (n+1/2)\Delta t$, so $\dot \theta (t)$ has to be calculated as the average
of $\dot \theta$ at the two adjacent time steps
\begin{align} \label{eq:lp1}
  \dot{\theta}(t) \approx \frac{1}{2}\big\{\dot\theta(t + \Delta t/2) +
  \dot \theta(t-\Delta t/2)\big\} .
\end{align}
The second order time derivative $\ddot \theta(t)$ is symmetrically
defined as
\begin{align}  \label{eq:lp2}
  \ddot \theta(t) \approx \frac{1}{\Delta t}\big\{\dot\theta(t + \Delta t/2) -
  \dot \theta(t-\Delta t/2)\big\}. 
\end{align}
The same applies for the twist variables $\mathbf\Delta$.
These definitions make the heat current $\bar J^Q_x(t)$ as defined
above naturally symmetric around $t$. An interesting aspect here is
that by choosing the time step as $\Delta t = 2Q^2$, the RCSJ
equations of motion discretized by the symmetric leap-frog scheme
actually reduce exactly to the RSJ equations of motion discretized
using an asymmetric forward Euler scheme.
Moreover $\Itot(t)$ becomes the sum of the super-, resistive, and
noise currents discretized by a forward Euler scheme as it should for
the RSJ model, while the voltages in \Eq{eq:heat-current} are kept
symmetric due to the definition \Eq{eq:lp1}.
The RSJ model is therefore best thought of as a special case of an
overdamped RCSJ model with $Q^2 = \Delta t/2$ (with our choice of
$\Delta t = 0.04$ this corresponds to $Q^2 = 0.02$).

We find that the heat current $\bar J^Q_x(t)$ is very sensitive to the
discretization used. In fact, it is critical to use the symmetric way
of defining $\bar J^Q_x(t)$ to obtain consistent results when
calculating the heat conductivity $\kappa$ either using a Kubo formula
or by applying a small temperature gradient.

\section{Results and discussion}\label{sec:results}

\subsection{Zero field thermal conductivity}

\begin{figure}[h!]
\includegraphics[width = 8cm]{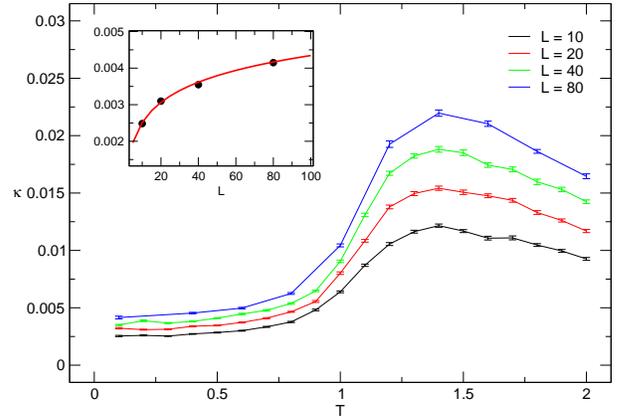}
\caption{(Color online) Heat conductivity $\kappa$ vs temperature $T$ 
for an $L \times L$ square lattice with underdamped RCSJ dynamics ($Q$ = 10). 
 The different curves correspond to different system sizes $L$.
 The inset shows the logarithmic dependence on system size $L$ at
  low $T$ ($T = 0.1$).
 The circles are simulation data and the smooth red curve is the analytic result 
 obtained from a linearized model.
\label{fig:rcsj_square_kappa@f0.00}
}
\end{figure}

Figure~\ref{fig:rcsj_square_kappa@f0.00} shows the thermal
conductivity $\kappa$ in zero magnetic field for fairly underdamped
RCSJ dynamics ($Q = 10$) for different system sizes $L$.  At low $T$
it tends to the spin wave value given by \Eq{eq:kappa_large_L}, which
in dimensionless units becomes $\kappa \sim (1/4\pi Q^2) \ln
(L/a)$. For large $Q$ this background value is quite small. For smaller values of
$Q$ the background increases and soon overwhelms the vortex
contribution. In the following it will therefore be subtracted.

\begin{figure}[h!]
\includegraphics[width = 8cm]{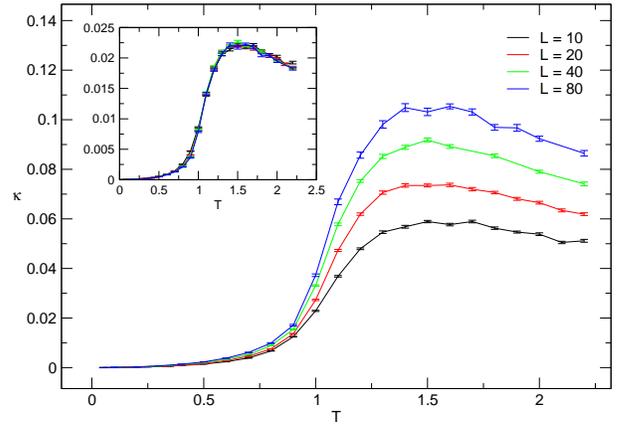}
\caption{(Color online) Heat conductivity $\kappa$ vs temperature
$T$ for an $L \times L$ square lattice with RSJ dynamics. Here the
  temperature independent spin wave background of $\kappa$ has been
  subtracted. The different curves correspond to different system
sizes $L$.  Inset: The
 same curves but scaled with $1/\ln(L/0.7)$.  The collapse is very
  good over the entire temperature range, except from close to the
  transition temperature $T_{\text{BKT}} \simeq 0.9$.
\label{fig:rsj_square_kappa@f0.00}
}
\end{figure}
The dependence on system size $L$ for low temperatures can be seen in
the inset of \Fig{fig:rcsj_square_kappa@f0.00}. The dependence is
logarithmic and follows very well the form in \Eq{eq:kappa_large_L},
shown as the red curve in the inset. In
\Fig{fig:rsj_square_kappa@f0.00} we can see $\kappa$ as a function of
temperature in the strongly overdamped limit $C \to 0$ (corresponding
to RSJ dynamics), but now with the harmonic spin wave background
subtracted. In RSJ dynamics where $C = 0$, the spin wave contribution
is formally proportional to $\delta(0)$, which translates to $1/\Delta
t$ in the numerics, where $\Delta t$ is the time step used in the
discretization (remember that RSJ dynamics with finite timestep
$\Delta t$ can be viewed as a special case of RCSJ dynamics with $Q^2
= \Delta t/2$). What is left after subtracting this part can be
interpreted as coming mainly from the motion of vortices.  The curves
of $\kappa$ start out very small for low temperatures, but increase
rapidly on approaching the
Berezinskii-Kosterlitz-Thouless~\cite{Berezinskii1,*Berezinskii2,*KT}
temperature $T_{\text{BKT}} \simeq 0.9$, where the unbinding of
thermally induced vortex-antivortex pairs makes a large contribution
to the thermal conductivity. At around $T = 1.4$ $\kappa$ reaches its
maximum value, followed by a slow decrease for higher
temperatures. Note that $\kappa$, even after subtracting the
background, shows a logarithmic dependence on the system size.  The
inset of \Fig{fig:rsj_square_kappa@f0.00} displays the curves for
different system sizes divided by the factor $\ln(L/a)$, with $a =
0.7$. The collapse of the curves onto a single one is very good over
the entire temperature range (except from very close to
$T_{\text{BKT}} \simeq 0.9$, where small deviations are expectedly
seen).

\begin{figure}[t]
\includegraphics[width = 8cm]{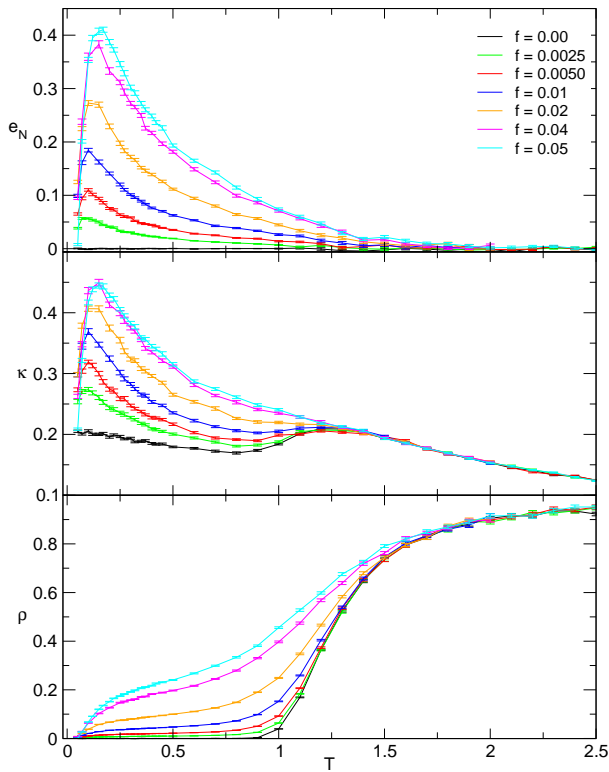}
\caption{(Color online) Nernst signal $e_N$, heat conductivity
  $\kappa$, and electrical resistivity $\rho$ vs temperature $T$ at
  different fillings $f$ for a 20$\times$20 square lattice with
  Langevin dynamics.
\label{fig:langevin_square_vs_T}
}
\end{figure}
\begin{figure}[t]
\includegraphics[width = 8cm]{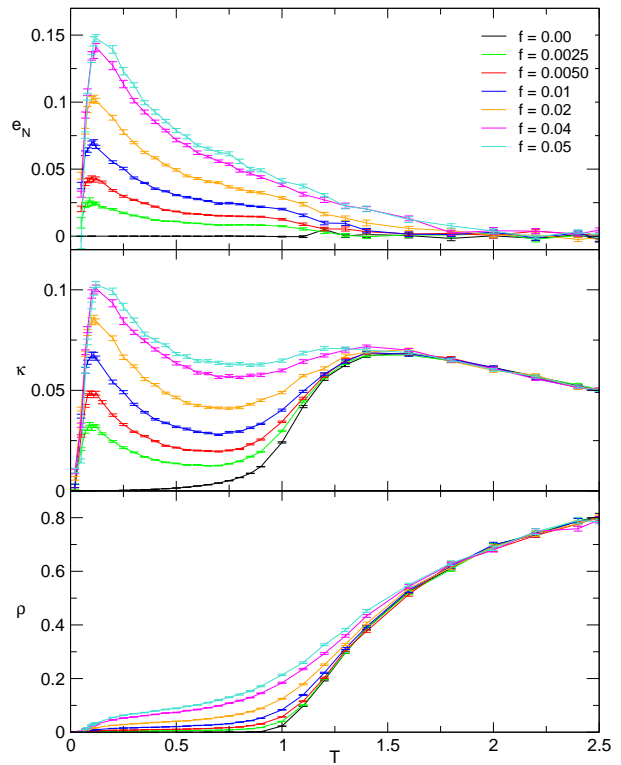}
\caption{(Color online) Nernst signal $e_N$, heat conductivity $\kappa$, and 
 electrical resistivity $\rho$ vs temperature $T$ at different 
  fillings $f$ for a 20$\times$20 square lattice with RCSJ dynamics ($Q$ = 0.5).
  In the $\kappa$ plot, the temperature independent spin wave background has been subtracted.
\label{fig:rcsj_square_vs_T}
}
\end{figure}

In the spin wave approximation the logarithmic size dependence seems to
be related to the long range Coulomb interation.
Screening can be introduced by adding capacitances to ground on
every grain, which removes the divergent behavior on length scales larger
than the screening length $\sqrt{C/C_0}$.
In a similar spirit, it seems likely that the logaritmic size
dependence of the \emph{vortex} contribution is tied to the unscreened
Coulomb interaction in the RCSJ model.

For Langevin dynamics, on the other hand, $\kappa$ is only weakly
dependent on system size and quickly converge to a size independent
bulk value (for $L \gtrsim 20$). Moreover, the finite size effects of
$e_N$ and $\rho$ are also negligible for $L \gtrsim 20$, for both
Langevin and RCSJ dynamics. We will therefore stick to lattices of
size $20\times20$ in the remainder of this paper.

At low $T$ for Langevin dynamics (see the lowest curve in the middle
panel of \Fig{fig:langevin_square_vs_T}) $\kappa$ goes to the spin
wave value \Eq{eq:kappa-langevin-linearized}, but decreases slightly
upon increasing the temperature until temperature induced vortices
become plentiful near the BKT transition, where $\kappa$ increases
again reaching a maximum around $T\approx 1.25$ and then starts to
decrease.

\subsection{Low fields}

We now turn to the case of a relatively weak applied transverse
magnetic field. In \Fig{fig:langevin_square_vs_T} and
\ref{fig:rcsj_square_vs_T} we see a collection of simulation results
for fillings $f = 0$ to $0.05$ on a $20\times20$ square lattice for
Langevin and RCSJ dynamics (Q = 0.5) respectively. The filling $f = B
\av{A_{\text{pl}}} / \Phi_0$, where $\av{A_{\text{pl}}}$ is the
average plaquette area, represents the average number of magnetic
field induced vortices per plaquette in the system. The lowest
nonzero filling $f = 0.0025 = 1/20^2$ corresponds to one field
induced vortex in the system.
Increasing the filling, i.e., raising the magnetic field will cause
the vortex density to increase, but as long as the typical vortex
separation is much larger than the coherence length $\xi$ ($\xi$ can
be thought of as the short distance cut-off or lattice spacing $a$ in
our model) the effects of discreteness are negligible and the model
should describe a continuous two-dimensional (or
quasi-two-dimensional) type-II superconductor. The case of strong
magnetic fields are discussed in Section~\ref{sec:highfield}.

Focusing on the Nernst signal $e_N$ in \Fig{fig:langevin_square_vs_T}
and \ref{fig:rcsj_square_vs_T}, we notice a very steep increase of
$e_N$ at low temperatures to a maximum between $T = 0.10$ and
$0.15$. At higher temperatures $e_N$ slowly decreases and the tail
persists up to about $T = 2$, which is roughly twice the BKT
transition temperature $T_\text{BKT} \approx 0.9$.  These features are
qualitatively similar for both Langevin and RCSJ dynamics. The main
difference between the models in the shape of $e_N$ is the
plateau-like part of the curves present in the RCSJ case around $T =
0.7$ for low fillings.  The peak height increases rapidly with
filling, before it starts to decrease for higher fillings.  The
position of the Nernst signal peak depends slightly on the filling $f$
and moves towards higher temperatures for larger fillings.

The qualitative features of the Nernst signal are in agreement with
experiments on several
superconductors~\cite{RiGross,WangOng2006,KokanovicCooper,PourretBehnia_NJP2009}.
A detailed comparison is, however, possible only if the temperature
and magnetic field dependences of the parameters $\gamma$, $R$, $I^c$
are taken into account.  For Josephson-junction arrays made up of
tunnel junctions the Ambegaokar-Baratoff
formula~\cite{AmbegaokarBaratoff1963a,*AmbegaokarBaratoff1963b} $I^c =
(\pi \Delta(T)/2eR) \tanh(\Delta(T)/2k_B T)$ can be used.  In bulk
superconductors $I^c$ is proportional to absolute square of the
superconducting order parameter $|\psi|^2$, which goes as $\sim
(T_c^\text{MF} - T)$, near $T_c^\text{MF}$.  The situation in the
high-$T_c$ cuprates is more complicated, since a model for the
relaxation rate, $\gamma(T)^{-1}$ in the Langevin case, or $R(T)$ in
the RCSJ case, is also required.  The thermoelectric coefficient
$\alpha_{xy} = e_N/\rho$ may have an advantage here, since both $e_N$
and $\rho$ are proportional to the relaxation rate, which therefore
drops out~\cite{MukerjeeHuse,Podolsky}. We will return to $\alpha_{xy}$ in the
next subsection.

In the second row of \Fig{fig:langevin_square_vs_T} and
\ref{fig:rcsj_square_vs_T} the heat conductivity $\kappa$ is plotted
as a function of temperature at different fillings $f$ (in the RCSJ
case the spin wave background has been subtracted). Note first how
similar the low temperature part of the curves of $\kappa$ are to the
Nernst signal. The onset and the peak positions of the two quantities
agree to a high degree.
For Langevin dynamics (\Fig{fig:langevin_square_vs_T}) $\kappa$ is
finite in the limit $T \to 0$.
For the two lowest fillings $f = 0.0025$ and $0.005$,
$\kappa$ increases quickly as a function of $T$ and then falls off
slowly below the $T \to 0$ value of $\sim 0.20$ to suddenly increase
again around $T_\BKT$ and reach a second maximum followed by slow a
decrease at higher temperatures. At fillings above $f = 0.005$ the
thermal conductivity follows the same pattern, but does not fall below
the $T \to 0$ value until temperatures above $T_\BKT$.  In the high
temperature regime $T \gtrsim 1.5$ all curves, regardless of filling
$f$, fall onto a single curve.
The curves for overdamped RCSJ dynamics with $Q=0.5$
(\Fig{fig:rcsj_square_vs_T}) share this feature of two maxima. In this
case, however, the falloff after the first maximum is even more
pronounced and persists up to higher fillings, at least $f =
0.05$. Note that the temperature independent background contribution
to $\kappa$ has been subtracted in this figure.

The double-peak behavior seems to indicate two separate contributions to the
thermal conductivity at high and low temperatures. The first maximum
at low $T$ is probably caused by the increased mobility of the field
induced vortices, which also gives the sharp rise of $e_N$.  This
contribution diminishes with increasing $T$, until the unbinding of
temperature induced vortex-antivortex pairs around $T_\BKT \simeq 0.9$
makes $\kappa$ large again. This latter contribution totally dominates
the previous one at higher temperatures, causing curves for different
fillings to converge.

Looking at the resistivity $\rho$, it displays, not the same but a
qualitatively similar $T$ dependence, for Langevin and RCSJ
dynamics. For Langevin dynamics the effect of varying filling $f$ is
somewhat more apparent, and the rise at $T_\BKT$ is also a bit
steeper than for RCSJ dynamics.

The results for the RCSJ model discussed previously have been for the
overdamped case $Q=0.5$. Upon reducing the damping and moving into the
underdamped regime, $e_N$ is effectively unchanged in the low
temperature region (apart from a trivial change of scale).  However,
$\kappa$ and $\rho$ change slightly at high temperatures $T \gtrsim
1.5$, in that $\kappa$ decays somewhat faster and the falloff starts
at a lower temperature, while $\rho$ increases more quickly as
function of $T$, than in the underdamped case.

\subsubsection{Vortex heat transport}

\begin{figure}[h!]
\includegraphics[width = 8cm]{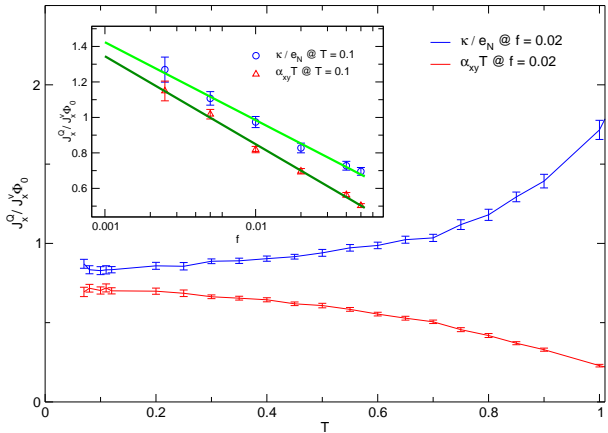}
\caption{(Color online) A plot of $\kappa /e_N$ and $\alpha_{xy} T$
  vs temperature $T$ for RCSJ dynamics ($Q=0.5$) on a square
  $20\times20$ lattice at filling $f=0.02$. Both of these measure the
  heat transport per vortex $(J_x^Q/J_x^v \Phi_0)$. Inset: The $f$ dependence
  of these two quantities at low T ($T$ = 0.1).  The two smooth green
  curves are fits to the form $a + b\ln f$ (light green: $a$ = 0.11,
  $b$ = -0.19; dark green: $a$ = -0.14, $b$ = -0.22).
\label{fig:alphaT_kappaovernernst}
}
\end{figure}

\begin{figure}[h!]
\includegraphics[width = 8cm]{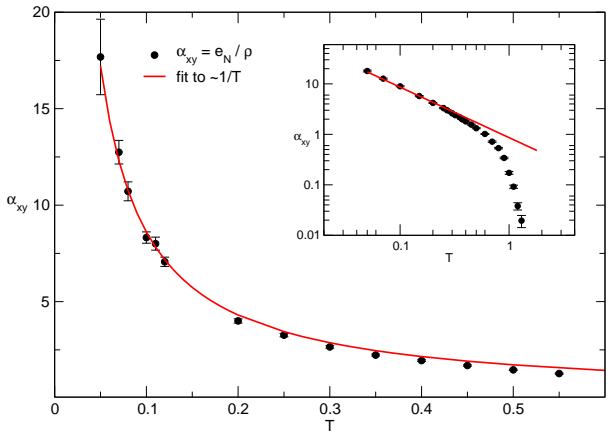}
\caption{(Color online) The off-diagonal component of the
  thermoelectric tensor $\alpha_{xy} = e_N / \rho$ vs temperature
  $T$ for RCSJ dynamics ($Q = 0.5$) on a square $20\times20$ lattice at
 filling $f=0.01$. At low $T$ the curve follows the power law $\sim 1/T$.
 In the high temperature region the falloff is much faster, about $\sim 1/T^8$.
\label{fig:alphaxy_vs_T_fit_rcsj}
}
\end{figure}
In our models there is no Hall effect ($\sigma_{xy} = 0$), leading to
$e_N = \alpha_{xy} / \sigma_{xx} = \alpha_{xy} \rho_{xx}$, and so we
can obtain yet another transport coefficient, the off-diagonal
component of the thermoelectric tensor $\alpha_{xy}$ from $e_N$ and
$\rho$. As mentioned one advantage with $\alpha_{xy}$ is that, unlike
$e_N$ and $\kappa$, it does not depend on the time constants $\gamma$
and $R$~\cite{MukerjeeHuse,Podolsky}.
This makes comparison with experimental data easier. Furthermore, from
phenomenological theories of vortex motion $\alpha_{xy}$ has the
interesting interpretation of the entropy per
vortex~\cite{HuebenerSeher}, suggesting that $\alpha_{xy} T$ can be
identified with the transported heat per vortex. This can also be seen
from the Onsager relation $\tilde \alpha_{xy} = \alpha_{xy} T$
together with the definition of the transverse electrothermal
conductivity $\tilde \alpha_{xy} = J^Q_x / E_y$, which is measured
under conditions with no applied temperature gradient. Now, since
$E_y$ is proportional to the transverse vortex current $J^v_x$ through
the relation $E_y = \Phi_0 J^v_x$, the quantity $J^Q_x / E_y$ is just
the ratio between the heat current and the vortex current, i.e., a
measure of the average transported heat per vortex.

Another way of calculating the heat transported per vortex, or more
precisely the ratio between the heat current and the vortex current,
is to simply divide the vortex contribution $\kappa$ of the
longitudinal heat conductivity (i.e., with the spin wave background
subtracted) with the Nernst signal, $\kappa / e_N =
(\frac{J^Q_x}{-\nabla_xT})/(\frac{E_y}{-\nabla_xT}) = J^Q_x / E_y$. Note,
however, that in this context the ratio $J^Q_x / E_y$ measures the
transported heat per vortex in a system driven by a temperature
gradient $\nabla_xT$, as opposed to the case of $\alpha_{xy} T = J^Q_x
/ E_y$, where the driving force is a transverse electric current
$J_y$. An equality of $\kappa / e_N$ and $\alpha_{xy} T$ would imply a
strict proportionality of the heat current $J^Q_x$ on the vortex
current $J^v_x$.

In \Fig{fig:alphaT_kappaovernernst} $\kappa / e_N$ and $\alpha_{xy} T$
from our simulations are plotted as functions of temperature
$T$. While they are not equal, they do agree very well at low $T$ (and
for low fillings $f$), so here one can approximately speak about
transported heat per vortex.  $\kappa / e_N$ is consistently larger
than $\alpha_{xy} T$, and the difference grows with increasing $T$
(and increasing $f$). A clue to why this happens can be found
considering the difference in what drives the vortex motion in the two
cases, as mentioned above. When applying a temperature gradient, heat
can be transported even without a net flow of vortices, being mediated
solely through the interactions between vortices at different
temperatures.  This is not possible when the driving force is the
transverse electric current, and naturally explains why $\kappa / e_N
$ is always greater than $\alpha_{xy} T$.
The magnetic field (or filling $f$) dependence of $\kappa / e_N$ and
$\alpha_{xy} T$ at a low fixed temperature is shown in the inset of
\Fig{fig:alphaT_kappaovernernst}. The plot is in lin-log scale and
clearly shows a logarithmic dependence at low fillings, which is quite
similar for both quantities.  Such a logarithmic dependence obtains
from an ideal gas treatment of the vortices, the Sackur-Tetrode
entropy per vortex being $\sim -\ln f$.  The vortices are, however,
strongly interacting. A crude way to estimate the interaction effects
on the transport entropy would be to assume that the available volume
per vortex is reduced by a factor of $N$, the number of vortices. This
then gives a contribution $\ln \Omega/N = -\ln f$ to the configurational
entropy, i.e., also a logarithmic dependence.

Also notice that in the low temperature region of
\Fig{fig:alphaT_kappaovernernst}, up to about $T \simeq 0.5$, $\kappa
/ e_N $ and $\alpha_{xy} T$ are only weakly temperature
dependent. This means that $\alpha_{xy}$ roughly falls of as $\sim
1/T$ for low temperatures. A fit to $\sim 1/T$ of $\alpha_{xy}$ for
RCSJ dynamics ($Q = 0.5$) at $f=0.01$ is displayed in
\Fig{fig:alphaxy_vs_T_fit_rcsj}. The inset in log-log scale reveals
that for temperatures above $T \simeq 0.8$, $\alpha_{xy}$ falls off
much faster, somewhere close to $\sim 1/T^8$.
These features are valid also for Langevin and RSJ dynamics,
as well as for other types of lattices.

\subsection{Intermediate and high fields - effects of granularity} \label{sec:highfield}

Going to higher magnetic fields the type of lattice structure starts
to play an important role, as geometric frustration will affect vortex
transport in this regime. The models at hand are then more valid as
descriptions of granular superconductors.
 
\subsubsection{Square lattices}
\begin{figure}[t!]
\includegraphics[width = 8cm]{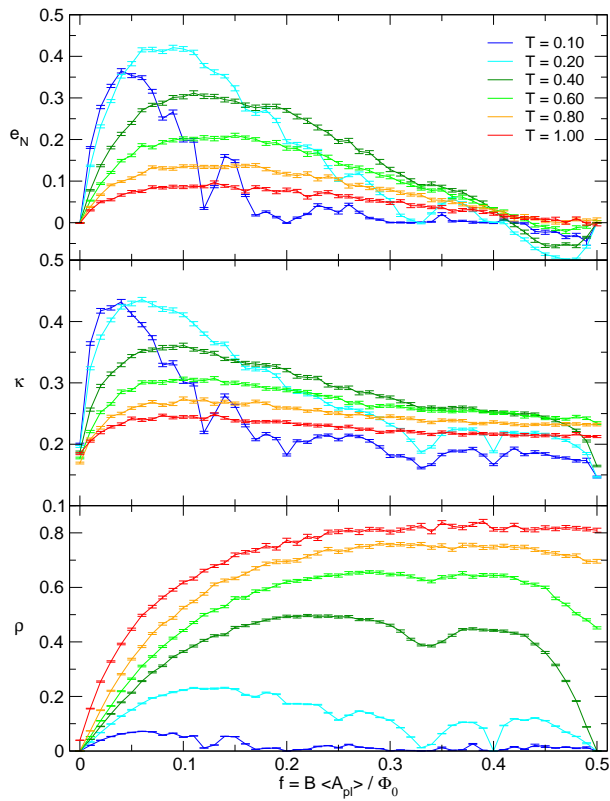}
\caption{(Color online) Nernst signal $e_N$, heat conductivity $\kappa$, and 
 electrical resistivity $\rho$ vs filling $f$ at different temperatures 
  for a 20$\times$20 square lattice with Langevin dynamics.
\label{fig:langevin_square_vs_f}
}
\end{figure}

\begin{figure}[t!]
\includegraphics[width = 8cm]{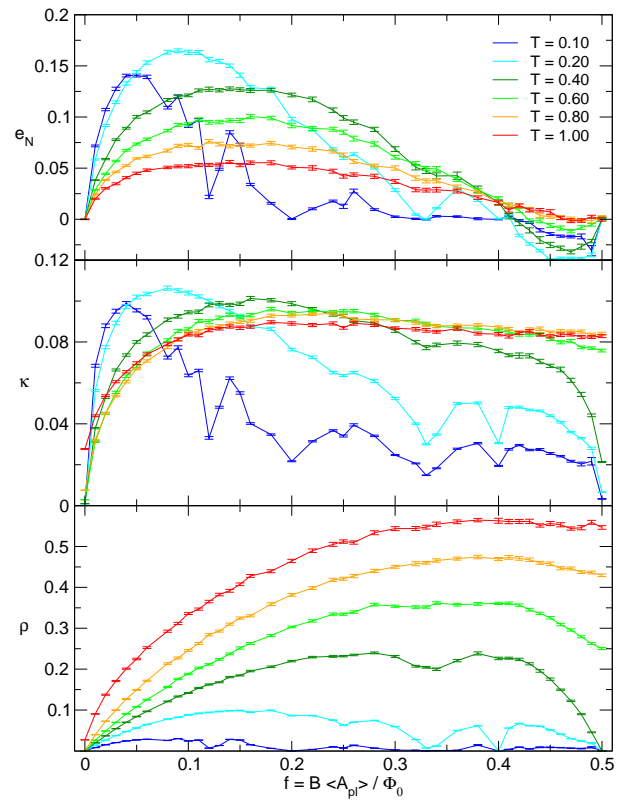}
\caption{(Color online) Nernst signal $e_N$, heat conductivity $\kappa$, and 
 electrical resistivity $\rho$ vs filling $f$ at different temperatures 
  for a 20$\times$20 square lattice with RSJ dynamics.
\label{fig:rsj_square_vs_f}
}
\end{figure}

Figures \ref{fig:langevin_square_vs_f} and \ref{fig:rsj_square_vs_f}
show simulation results as a function of filling $f$ for Langevin and
RSJ dynamics (corresponds to RCSJ dynamics with $Q=\sqrt{0.02}$) on a
$20\times20$ square lattice.  At low fillings the Nernst signal $e_N$
and the heat conductivity $\kappa$ both show a sharp increase
culminating in a maximum around $f=0.05-0.15$, depending on
temperature, followed by a decrease up to half-filling. This
``tilted-hill'' profile of the Nernst signal seems to be generic
and is found in a number of experiments on cuprates and ordinary
type-II
superconductors~\cite{Xu2000,*WangOng2006,PourretBehnia_PhysRevB2007}.
Note how $e_N$ is always zero at $f = 0$ and $1/2$, due to vortex -
vacancy symmetry.  The heat conductivity $\kappa$ on the other hand
stays finite at $f=1/2$.  On a perfectly periodic lattice all physical
quantities are, because of the symmetry of the XY model Hamiltonian
[\Eq{eq:XY}], periodic in filling $f$, with period one, and also
mirror symmetric (for $\rho$, $\kappa$) or antisymmetric ($e_N$)
around $f=1/2$. Thus, all information is contained in the region $f =
0 \to 0.5$, which is displayed here.

\begin{figure}[]
\includegraphics[width = 7.7cm]{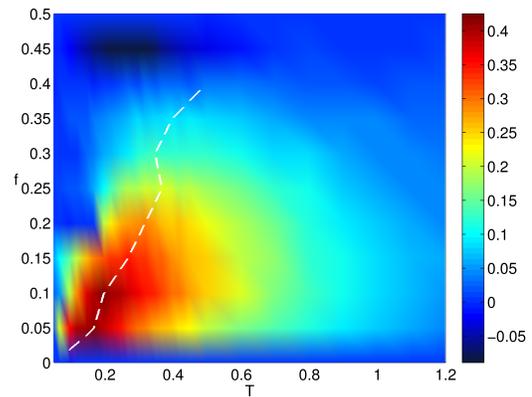}
\caption{(Color online) Contour plot of the Nernst signal $e_N$ for a
  20$\times$20 lattice with Langevin dynamics in the temperature -
 filling plane. The dashed white line joins the maxima of 
  $e_N$.
  \label{fig:langevin_square_nernst_contour}
}
\end{figure}

\begin{figure}[t]
\includegraphics[width = 8cm]{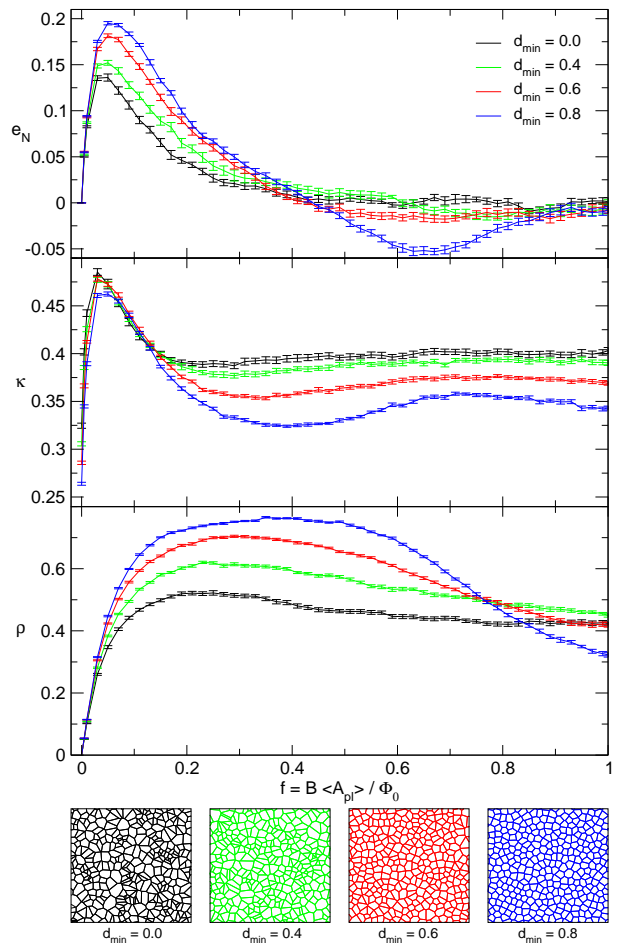}
\caption{(Color online) Nernst signal $e_N$, heat conductivity $\kappa$, and 
electrical resistivity $\rho$ vs filling $f$ at $T = 1$ for
random lattices of size $20 \times 20$
with different degrees of disorder set by the parameter $d_{\text{min}}$ using Langevin dynamics. 
Each curve is an average over 16 disorder realizations.
The lower panel shows examples of Voronoi lattices obtained from random
packings of grains with different
$d_\text{min}$.
\label{fig:langevin_random_vs_f}
}
\end{figure}

Now, lowering the temperature the curves have significant structure
due to geometric frustration as the filling is varied through
different commensurate values. At fillings such as $f$ = 1/8, 1/5,
1/3, 2/5, all three transport coefficients $e_N$, $\kappa$ and $\rho$
are reduced due to vortex pinning to the underlying lattice. This is
particularly apparent at the 2nd lowest temperature $T = 0.2$ (cyan
colored curves) at f = 1/3 and 2/5. The observant reader may also have
noticed that the Nernst signal actually goes negative in a region
below half-filling.  In fact a small region of negative Nernst signal
appear also right below $f=1/3$, and it is plausible that this occurs
below other commensurate fillings as well, over certain temperature
intervals.
This sign reversal of $e_N$ is a new effect~\cite{AnderssonLidmar}
seen in all our simulations independent of the type of dynamics used
(Langevin or over-/underdamped RCSJ) and persists also for lattices
with moderate geometric disorder (see
e.g. \Fig{fig:langevin_random_vs_f}). As discussed in a previously
published paper of ours~\cite{AnderssonLidmar}, a negative vortex
Nernst signal [given the definition in \Eq{eq:nernst_def}] implies
vortex transport in the direction opposite of heat transport. We argue
that this is possible, since around these special fillings there is a
temperature regime, where mobile vortex vacancies can exist on top of
a pinned vortex lattice. The vortex vacancies then diffuse down the
applied temperature gradient, creating a net vortex flow in the
opposite direction and thus a negative Nernst signal.

\begin{figure}[t]
\includegraphics[width = 8cm]{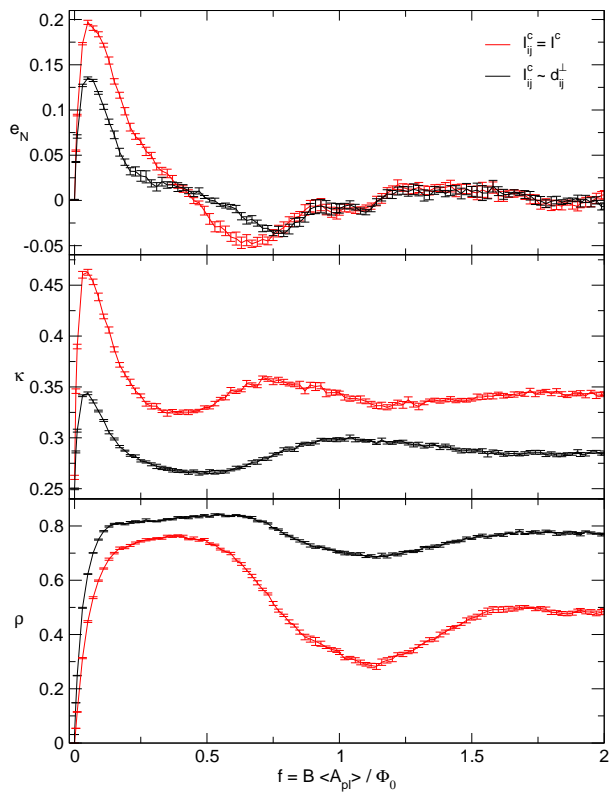}
\caption{(Color online) Nernst signal $e_N$, heat conductivity $\kappa$ 
  and electrical resistivity $\rho$ vs filling $f$ at $T=1$ for
  random lattices of size $20 \times 20$ 
  with $d_\text{min} = 0.8$, using Langevin dynamics.
  Results shown are for two models with different 
  critical current distributions $I^c_{ij} = I^c$ (red) and $I^c_{ij} \sim d^\perp_{ij}$ (black). 
  Each curve is an average over eight disorder realizations.
\label{fig:langevin_random_varcoupling_vs_f}
}
\end{figure}

Comparing the Nernst signal versus $f$ for Langevin and RSJ dynamics
in \Fig{fig:langevin_square_vs_f} and \ref{fig:rsj_square_vs_f}
respectively, one sees an almost exact agreement (as opposed to $e_N$
versus $T$ at low fillings, where Langevin and RSJ dynamics are less
similar). Also increasing the damping parameter $Q$ for RCSJ dynamics,
i.e., going from the overdamped to the underdamped limit results in
hardly any changes in the qualitative filling dependence of $e_N$,
$\kappa$ and $\rho$. This should indicate that these quantities in
this regime are governed by geometric frustration effects and are
rather insensitive to model specific details.

As a summary of the Nernst effect on a square lattice (for Langevin
dynamics) we provide in \Fig{fig:langevin_square_nernst_contour} a
contour plot of $e_N$ in the temperature - filling plane. The red
regions indicate a large Nernst signal and the blue ones a signal
which is close to zero, or even negative (the dark blue blob in the
upper left corner).

\subsubsection{Random lattices}

\begin{figure}[]
\includegraphics[width = 8cm]{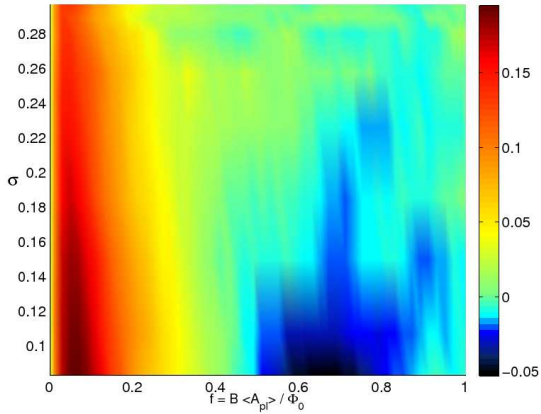}
\caption{(Color online) Contour plot of the Nernst signal $e_N$ 
in the filling - grain size standard deviation plane for random 
lattices with Langevin dynamics at $T = 1$.
\label{fig:langevin_random_contour}
}
\end{figure}

In order to model random granular superconductors we have carried out
simulations on randomly connected networks as defined in
Sec.~\ref{sec:lattice-structure}.
In \Fig{fig:langevin_random_varcoupling_vs_f} we compare $e_N$,
$\kappa$ and $\rho$ as functions of filling at $T = 1$ for two
different models with Langevin dynamics on a random lattice with
$d_{\text{min}} = 0.8$. In the first model (red curve) we set the
critical currents of every junction to a constant $I^c_{ij} =
I^c$. The second model (black curve) has the critical currents
proportional to the contact area (or strictly contact length
$d^\perp_{ij}$ in 2D) between the grains, $I^c_{ij} \sim
d^\perp_{ij}$, see \Fig{fig:dperp_rji}.

In a geometrically disordered system without perfect periodicity the
Nernst signal and the resistivity are no longer periodic as a function
of filling. We have therefore extended the curves up to $f =
B\av{A_\text{pl}}/\Phi_0 = 2$ ($\av{A_\text{pl}} = 1/2$ for the random
lattices). The Nernst signal shows an even steeper increase at low $f$
and peaks even earlier than in the square lattice case. We also see
that $e_N$ is somewhat reduced in the model with $I^c_{ij} \sim
d^\perp_{ij}$. The negative region is narrower but above $f = 1$ the
curves are essentially the same. Looking at the heat conductivity the
two models display quite similar behavior as a function of
temperature, although in the second model the amplitude of $\kappa$ is
effectively halved. A more dramatic difference can be seen in $\rho$
though, where the model with added disorder of the critical currents
seems almost like a smoothed-out version of the first one, for which
$\rho$ has more dramatic features at high fillings.
At very high fillings $f \gg 1$ the flux through each plaquette modulo
$\Phi_0$ becomes approximately random in $[0,\Phi_0]$ making the Nernst
signal vanish and $\kappa$ and $\rho$ constant.
The results for RCSJ dynamics show a qualitatively similar behavior,
with some minor quantitative differences.

Results for different strengths of randomness (different values of
$d_{\text{min}}$, see Sec.~\ref{sec:lattice-structure}) are shown in
\Fig{fig:langevin_random_vs_f}. From this figure it is also apparent
that much of the structure in $e_N$, $\kappa$ and $\rho$ as a function
of filling is reduced as we move from less towards more disordered
lattices.  (The apparent lack of visible geometric frustration effects
at specific fillings are, however, in the case of these random
lattices due to the fact that we average over many (8 or 16) different
disorder realizations.)

\Fig{fig:langevin_random_contour} summarizes the Nernst effect in our
random lattices. It displays a contour map of the Nernst signal in the
filling - grain size standard deviation plane. The large blue region
at the bottom corresponds to a negative $e_N$, whereas the red region
to the left represents a large positive Nernst signal.

\section{Conclusions}

We have modeled heat and charge transport in two-dimensional granular
superconductors.  We have considered regular square arrays and
randomly connected networks.  Square arrays can be artificially
fabricated using lithography, while random arrays of varying degrees
of disorder may occur naturally in inhomogeneous thin-film
superconductors.  We consider two models for the dynamics of the
superconductors, relaxational Langevin dynamics and RCSJ dynamics.  An
expression for the heat current is derived from these models.
For the Langevin dynamics the heat current expression
\eqref{eq:heat-current} is consistent with previous expressions
derived microscopically or from time-dependent Ginzburg-Landau
theory~\cite{Schmid,CaroliMaki,Hu}.  For the RCSJ model, however, it
differs in that it contains the total current, including the
capacitative and resistive currents, which shunt the supercurrent
through the junction.
The expressions are used in numerical simulations to calculate the
heat conductivity, the resistivity, and the Nernst signal. We find an
anomalous logarithmic size dependence in the heat conductivity for the
RCSJ model in zero magnetic field.  This type of dependence is present
also in the spin wave approximation valid at low $T$.  We also find
$\kappa$ to be divergent in the limit when the shunting capacitor goes
to zero, showing that the RSJ model without capacitors is pathological
from the point of view of heat conduction.
The Nernst signal and the resistivity are still well behaved in this limit.

From our numerical simulations, we further find a highly nontrivial
nonmonotonous temperature dependence in $\kappa$ at low magnetic
fields in both models.
In this regime granularity appears to have a negligible influence on
the transport properties, and our results should apply also to
two-dimensional phase-fluctuating bulk superconductors.  For low
temperature $T \ll T_\text{BKT} \approx 0.9$ and magnetic fields it is
possible to define the transported heat per vortex, which is found to
depend logarithmically on filling $f$, while being approximately
temperature independent.  Note, however, that in our phase-only models
the vortices are coreless.  At higher $T$ thermally excited vortices
and antivortices start to influnce the results and dominate the
response.

At higher fields, granularity becomes important and geometric
frustration strongly influences the Nernst signal, heat conductivity,
and resistivity, leading to a highly intricate magnetic field
dependence as shown in Figs.~\ref{fig:langevin_square_vs_f}
--\ref{fig:langevin_random_contour}. These signatures should be
possible to obtain directly in experiments on regular Josephson-
junction arrays, similar to those carried out for the resistivity in
Ref.~\onlinecite{Baek}, or in patterned thin film superconductors.

\begin{acknowledgments}
Support from the Swedish Research Council (VR) and
Parallelldatorcentrum (PDC) is gratefully acknowledged.  
\end{acknowledgments}

%\bibliography{heat}
%Merlin.mbs v4.21 2009-07-09.
%

\end{document}